\documentclass{aa}  

\usepackage{graphicx}
\usepackage{txfonts}
\usepackage{natbib}
\bibpunct{(}{)}{;}{a}{}{,}
\usepackage[colorlinks=true, citecolor=blue]{hyperref}
\usepackage{multirow}
\usepackage{array,booktabs}
\begin{document}

   \title{Low-frequency monitoring of flare star binary CR~Draconis: Long-term electron-cyclotron maser emission}
   \titlerunning{Low-frequency monitoring of flare star binary CR~Draconis}


   \author{J.~R.~Callingham\inst{1,2}
          \and
          B.~J.~S.~Pope\inst{3,4,5}
          \and
            A.~D.~Feinstein\inst{6,7}
          \and
            H.~K.~Vedantham\inst{2,8}
         \and
            T.~W.~Shimwell\inst{2,1} 
          \and
            P.~Zarka\inst{9}
          \and
            C.~Tasse\inst{10,11}
          \and
          L.~Lamy\inst{9,12}
          \and
            K.~Veken\inst{1,2}
          \and
            S.~Toet\inst{1,2}
            \and
            J.~Sabater\inst{13}
            \and
            P.~N.~Best\inst{13}
            \and
            R.~J.~van Weeren\inst{1}
            \and
            H.~J.~A.~R\"{o}ttgering\inst{1}
            \and
            T.~P.~Ray\inst{14,15}
          }

   \institute{Leiden Observatory, Leiden University, PO Box 9513, 2300 RA, Leiden, The Netherlands\\
        \email{jcal@strw.leidenuniv.nl}
        \and
        ASTRON, Netherlands Institute for Radio Astronomy, Oude Hoogeveensedijk 4, Dwingeloo, 7991 PD, The Netherlands
        \and
             Center for Cosmology and Particle Physics, Department of Physics, New York University, 726 Broadway, New York, NY 10003, USA
        \and
             Center for Data Science, New York University, 60 5th Ave, New York, NY 10011, USA
        \and
             NASA Sagan Fellow
        \and
             Department of Astronomy and Astrophysics, University of Chicago, 5640 S. Ellis Ave, Chicago, IL 60637, USA
        \and 
             NSF Graduate Research Fellow
        \and
             Kapteyn Astronomical Institute, University of Groningen, PO Box 72, 97200 AB, Groningen, The Netherlands
        \and
             LESIA, Observatoire de Paris, Universit\'{e} PSL, CNRS, Sorbonne Universit\'{e}, Universit\'{e} de Paris,  5 Place Jules Janssen, 92195 Meudon, France
        \and
             GEPI, Observatoire de Paris, CNRS, Universite Paris Diderot, 5 place Jules Janssen, 92190 Meudon, France
        \and
             Department of Physics \& Electronics, Rhodes University, PO Box 94, Grahamstown, 6140, South Africa
        \and
             LAM, Aix-Marseille Universit\'{e}, CNRS, 38 Rue Fr\'{e}d\'{e}ric Joliot Curie, 13013 Marseille, France
        \and 
             SUPA, Institute for Astronomy, Royal Observatory, Blackford Hill, Edinburgh, EH9 3HJ, UK
        \and
            Dublin Institute for Advanced Studies, Astronomy \& Astrophysics Section, 31 Fitzwilliam Place, Dublin 2, Ireland
        \and
            School of Physics, Trinity College, Dublin 2, Ireland
             }

   \date{Received 10 August 2020 / Accepted 6 February 2021}

\abstract{Recently detected coherent low-frequency radio emission from M dwarf systems shares phenomenological similarities with emission produced by magnetospheric processes from the gas giant planets of our Solar System. Such beamed electron-cyclotron maser emission can be driven by a star-planet interaction or a breakdown in co-rotation between a rotating plasma disk and a stellar magnetosphere. Both models suggest that the radio emission could be periodic. Here we present the longest low-frequency interferometric monitoring campaign of an M dwarf system, composed of twenty-one $\approx$8\,hour epochs taken in two series of observing blocks separated by a year. We achieved a total on-source time of 6.5\,days. We show that the M dwarf binary CR\,Draconis has a low-frequency 3$\sigma$ detection rate of 90$^{+5}_{-8}$\% when a noise floor of $\approx$0.1\,mJy is reached, with a median flux density of 0.92\,mJy, consistent circularly polarised handedness, and a median circularly polarised fraction of 66\%. We resolve three bright radio bursts in dynamic spectra, revealing the brightest is elliptically polarised, confined to 4\,MHz of bandwidth centred on 170\,MHz, and reaches a flux density of 205\,mJy. The burst structure is mottled, indicating it consists of unresolved sub-bursts. Such a structure shares a striking resemblance with the low-frequency emission from Jupiter. We suggest the near-constant detection of high brightness temperature, highly-circularly-polarised radiation that has a consistent circular polarisation handedness implies the emission is produced via the electron-cyclotron maser instability. Optical photometric data reveal the system has a rotation period of 1.984$\pm$0.003\,days. We observe no periodicity in the radio data, but the sampling of our radio observations produces a window function that would hide the near two-day signal. 
}

\maketitle

%

\section{Introduction}

Low-frequency ($\lesssim$ 200\,MHz) emission from stellar systems encode the conditions of the outer stellar corona \citep{2017ApJ...836L..30L,2018ApJ...856...39C,2019ApJ...871..214V,2020NatAs.tmp...34V}. For example, low-frequency emission can be used to recover the kinematic properties of massive plasma releases that accompany magnetic eruptive events in the outer stellar corona, such as coronal mass ejections (CMEs) \citep{1950AuSRA...3..387W,1985srph.book.....M,1985ARA&A..23..169D}. Understanding the properties of CMEs is considered especially pertinent for evaluating the space weather experienced by planets around other stars \citep{2015A&A...580A..80K,2016Ap&SS.361..253B}. Additionally, detecting the low-frequency emission expected from a star-planet interaction, as modelled off a scaled-up Jupiter-Io system, provides a direct measurement of the orbital period of a putative exoplanet, the approximate size of the exoplanet, magnetic field alignment and strength, and the Poynting flux incident on the exoplanet \citep{2007P&SS...55..598Z,2011A&A...531A..29H,2016pmf..rept.....L,2020NatAs.tmp...34V}.
 
While previous searches for radio-bright stellar systems have predominately been conducted at gigahertz frequencies \citep[e.g.][]{1989ApJS...71..895W,1999AJ....117.1568H}, a renaissance in observing stellar systems at low frequencies is occurring due to the maturation of the Murchison Widefield Array \citep[MWA;][]{Tingay2013}, the Giant Metrewave Radio Telescope \citep[GMRT;][]{1991ASPC...19..376S}, and the LOw-Frequency ARray \citep[LOFAR;][]{vanHaarlem2013}. In particular, the LOFAR Two-metre Sky Survey \citep[LoTSS;][]{2019A&A...622A...1S} has demonstrated that it is now routine to achieve $\approx$100\,$\mu Jy$ root-mean-square noise at $\approx$150\,MHz. Since the brightness of the low-frequency emission from nearby star-planet interactions and stellar magnetic processes is expected to be $\lesssim$1\,mJy \citep{2018MNRAS.478.2835L,2018MNRAS.478.1763L,2019MNRAS.484..648P,2020NatAs.tmp...34V}, LoTSS provides an unparalleled dataset to blindly search for radio-bright stellar systems \citep{2019RNAAS...3...37C}. 

Significant advances are also occurring in optical photometry with the operation of the Transiting Exoplanet Survey Satellite \citep[TESS;][]{2015JATIS...1a4003R}. TESS is conducting accurate photometric observations of over 200,000 stars with a two minute cadence, facilitating not only a comprehensive search for transiting exoplanets but also for stellar flares and rotation periods \citep{2016ApJ...829...23D,Gunther2019}. Preferentially, optical stellar flares appear to occur on fast rotating, young M dwarfs \citep{Gunther2019,feinstein20}, which are $\sim$0.1 to 0.6\,M$_{\odot}$ and the most common type of stars in our Milky Way \citep{2006AJ....132.2360H}. While it is possible that these are connected to CMEs, no conclusive association with a solar CME signature such as a Type II radio burst has been conclusively made \citep{osten17,villadsen17,alvarado20}. The impact that M dwarf stellar flares have on their surrounding exoplanets is highly contested, particularly because it is unclear whether CMEs accompany the flares \citep{Gunther2019}.

Recently, \citet{CallinghamPopulation} used LoTSS to identify a population of M dwarfs that displayed highly circularly polarised ($\gtrsim$\,60\%), broadband, high brightness temperature ($>$\,10$^{12}$\,K) low-frequency emission. The population is composed of M dwarfs from across the main sequence (M1 to M6) and at all chromospheric activity levels. Quiescent, slow-rotating ($\gtrsim$\,2\,d) M dwarfs were as likely to be detected as active, fast rotating stars. Such results are in contrast to stellar systems identified as radio bright at gigahertz frequencies, which preferentially tend to be fast rotating chromospherically and coronally active late M dwarfs \citep[e.g.][]{2012ApJ...746...23M,2018ApJ...856...39C,2019ApJ...871..214V,2019NatAs...3...82C}. Additionally, previous highly circularly polarised emission detected from ultracool and M dwarfs at gigahertz frequencies appear bursty in nature, generally lasting $\lesssim$2 hours \citep[e.g.][]{2008ApJ...684..644H,2018ApJ...854....7L,Zic2019,2019ApJ...871..214V,Zic2020}. In comparison, the coherent emission detected by \citet{CallinghamPopulation} was minimally variable and long duration ($\gtrsim$8\,hour). It is possible that a common emission mechanism acts in both high- and low-frequency selected samples, with fast-rotating late M dwarfs more likely to have the necessary high magnetic field strengths to produce coherent radio emission at gigahertz frequencies.

The proposed acceleration mechanisms that could produce the observed coherent low-frequency emission from most of the LoTSS detected M dwarfs are similar to the magnetospheric processes observed on the Solar System's gas giant planets and ultracool dwarfs \citep{2008ApJ...684..644H,vedanthambrowndwarf,CallinghamPopulation}. For example, the star itself can produce Jovian-like emission through electric field-aligned currents resulting from a breakdown in co-rotation between a rotating plasma disk and its magnetosphere \citep{schrijver09,2012ApJ...760...59N,2017ApJ...846...75P}. It is also possible that an exoplanet in orbit around a star could produce low-frequency coherent emission through the relative motion of a planet through a star's magnetosphere \citep{2004ApJ...612..511L,2007P&SS...55..598Z,2018ApJ...854...72T,2019MNRAS.488..633V,2020NatAs.tmp...34V,Mahadevan2021}. 

In both incidences the radio emission is produced via the electron-cyclotron maser instability (ECMI) mechanism in magnetised, plasma-depleted regions \citep{1979ApJ...230..621W,2006A&ARv..13..229T}. Alternatively, especially for the most active M dwarfs in the \citet{CallinghamPopulation} sample, it is possible that fundamental plasma emission could stochastically produce the observed coherent low-frequency emission \citep{2001A&A...374.1072S}.

One prediction of the breakdown of co-rotation and star-planet interaction models is that the radio emission could be periodic. This is because the ECMI radiation will be beamed along the surface of a cone that is modulated relative to our line of sight by either the rotation of the star or the orbit of a satellite. Such modulations are strongly dependent on the viewing geometry, and on the alignment of the magnetic and rotation/orbital axes. Unfortunately, the standard sampling of LoTSS is too sparse to be sensitive to the expected periodicities of $\sim$0.5 to 10 days \citep{2020NatAs.tmp...34V,CallinghamPopulation}, with most of the M dwarf detections only having two to three eight hour LoTSS exposures. Additionally, the detection strategy of searching images that were synthesised over the entire LoTSS observing time is biased towards detecting sources whose emission is aligned with our line of sight \citep{CallinghamPopulation}.

Fortunately, one of the stars from the \citet{CallinghamPopulation} sample is coincidentally located in a LoTSS deep-field \citep{Pepe2020} -- the European Large-Area ISO Survey-North 1 \citep[ELAIS-N1;][]{2000MNRAS.316..749O} field. As part of the LoTSS Deep Field data release 1 \citep{Pepe2020}, the star has 21 independent $\approx$8\,hour 146\,MHz exposures taken over two years, serendipitously providing a comprehensive dataset to test for any periodicity in the radio time series and to study the long-term evolution of coherent low-frequency emission.  

The low-frequency emitting star system that is located in the ELAIS-N1 field is CR\,Draconis (BD+55\,1823; GJ\,9552; hereafter CR\,Dra). CR\,Dra is a young M1.5Ve dwarf binary that has been proposed to have a period of $\approx$\,4\,yr \citep{2008AJ....136..974T}, but there are conflicting measurements for its orbit \citep{shkolnik2010,sperauskas19}. At least one of the two M dwarf components is a flare star, with an observed B-band ($\approx$450\,nm) flare rate of $\approx$1 flare per 10 hours \citep{2018JAVSO..46...21V}. CR\,Dra has not been previously detected in the radio despite previous sensitive searches at frequencies between 325\,MHz and 1.4\,GHz \citep{2007ApJ...666..201T,2008MNRAS.383...75G,2009MNRAS.395..269S}. CR\,Dra is also coronally and chromospherically active. It has a soft 0.2 to 2\,keV X-ray luminosity of 3.7$\times$10$^{29}$\,erg\,s$^{-1}$ \citep{2016A&A...588A.103B,CallinghamPopulation}, implying a coronal temperature as high as $\sim$7\,MK \citep{2015A&A...578A.129J}. Due to this hot corona, CR\,Dra is also one of the few stars identified by \citet{CallinghamPopulation} in which the observed low-frequency radiation could be generated by fundamental plasma emission \citep{2001A&A...374.1072S}. 

In this paper, we use the LOFAR ELAIS-N1 deep-field and TESS datasets on CR\,Dra to: 1) determine the optical flare statistics and photometric rotation period; 2) study the long term evolution of the coherent highly-circularly polarised low-frequency emission; 3) conduct a search for periodicity in low-frequency radio emission, and; 4) test if the properties of the low-frequency emission is consistent with plasma or ECMI radiation.

\begin{figure*}
\begin{center}
\includegraphics[scale=0.4]{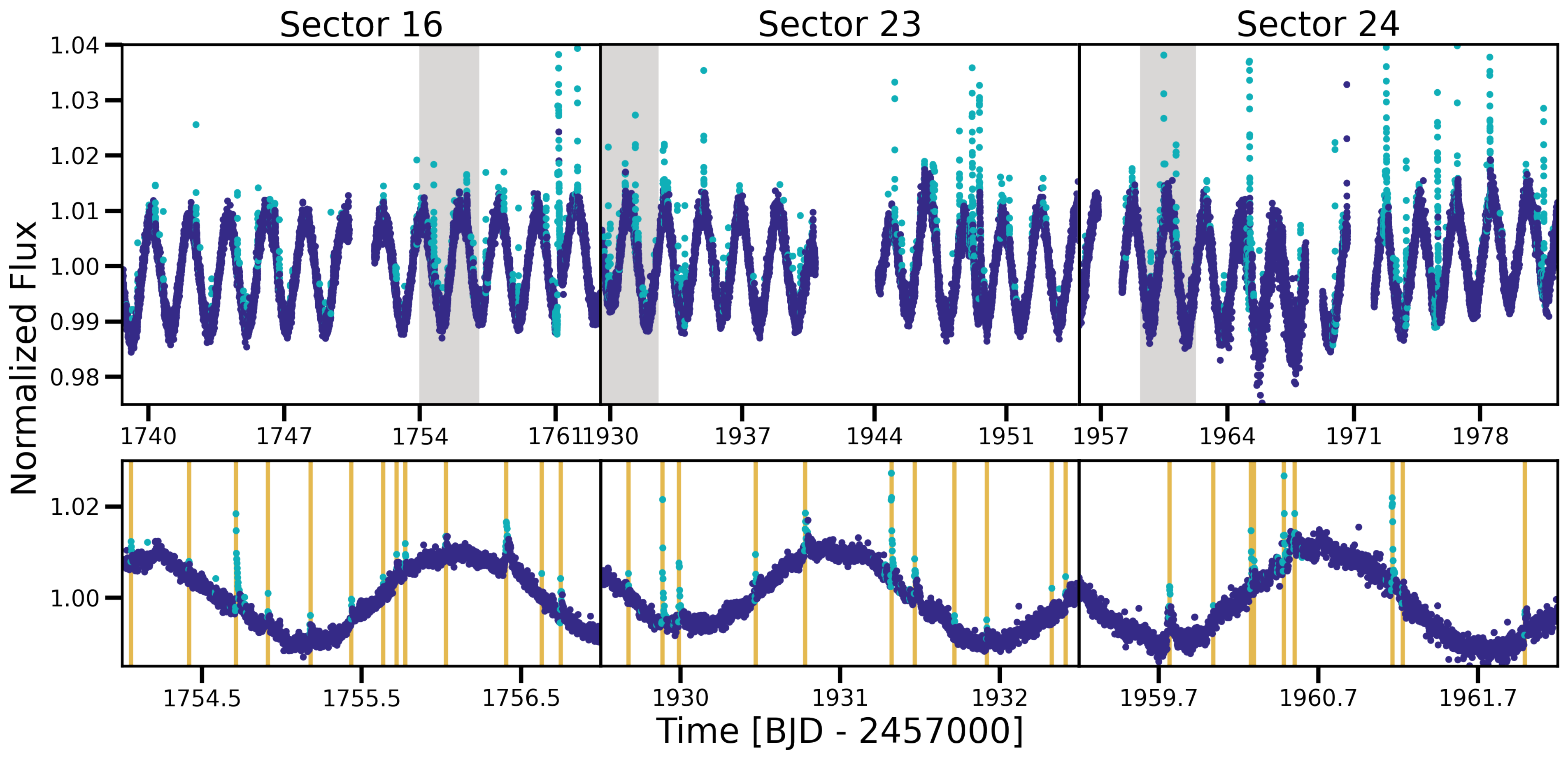}
 \caption{Lightcurves of CR~Dra in all three TESS Sectors for which it was observed. Top panels: total lightcurves from each sector. Bottom panels: zoom-in on the shaded regions within each sector. Teal represents cadences with \texttt{stella}-assigned probabilities above our threshold of 0.6; purple represents cadences below this probability. Flares found by \texttt{stella} are marked by the orange vertical lines. The vertical scale has been clipped to better show the rotational variability; the highest amplitude flare is nearly a factor of two in amplitude, as illustrated in Figure~\ref{fig:brightest_flares}.}
\label{fig:all_flares}
\end{center}
\end{figure*}

\section{TESS observations, flares, and photometric rotation period}
\label{sec:flares_prot}

CR~Dra has been observed with TESS in Sectors 16 (2019-09-11 to 2019-10-07), 23 (2020-03-18 to 2020-04-16), 24 (2020-04-16 to 2020-05-13), and 25 (2020-05-13 to 2020-06-08), in the TESS Input Catalog as TIC\,207436278 \citep{tic}. Processed data are publicly available for Sectors 16, 23, and 24 at the time of writing. 

We retrieved data from the Mikulski Archive for Space Telescopes for the available sectors, and pre-processed these data using \texttt{lightkurve} \citep{lightkurve}. We inspected the default pixel masks in the target pixel files and found them to be satisfactory. We then proceeded to initially reduce the standard Pre-search Data Conditioning Simple Aperture Photometry (PDC-SAP) 2-min cadence light curve products by removing invalid values and normalising each light curve to unity. 

The TESS light curves reveal significant variability in the form of rotational modulation and a large number of stellar flares. Previous flare-detection methods have relied on light curve detrending before applying outlier heuristics. However, high-amplitude variability, such as that present in these light curves, can prove challenging to completely remove and can lead to over-fitting, thereby removing low-amplitude flares. 

To overcome this bias, we applied the convolutional neural network (CNN) $\texttt{stella}$ \citep{feinstein20_joss}, trained on the \citet{Gunther2019} M dwarf flare catalogue compiled from TESS Sectors 1-2. The \texttt{stella} CNNs provide a probability for every cadence, or each individual data point, that it is part of a flare \citep{feinstein20}. For example, a cadence with a probability of 0.6 means the trained CNN models believe with 60\% confidence this cadence is part of a true flare. We used the ten pre-trained models as applied by \cite{feinstein20}\footnote{Models are hosted on MAST: \url{https://archive.stsci.edu/hlsp/stella}.} and averaged the probabilities for our final analysis. We selected a threshold value = 0.6, as used in \cite{feinstein20}, and weighted our analysis by probability; flares with higher probability values are weighted more heavily. Due to the small number of light curves in this study, we inspected each flare. Even though the CNN models assigned some of the flares a 60\% probability of being true flares, we found each to demonstrate true flare-like behaviour.

The results of this analysis are displayed in Figure~\ref{fig:all_flares}, where the light curves are coloured by the probability that each time sample is part of a flare. The bottom panel zooms in on a 3-day region in each TESS Sector, with identified flares highlighted by orange vertical lines. No systematics-correction detrending was applied to these light curves, as the flare-detecting neural network is trained on data with uncorrected systematics and has learned to filter these. The \texttt{stella} algorithm includes additional checks for classifying an event as a flare, as the CNN was trained on a catalogue which is incomplete at low flare amplitudes \citep{feinstein20}.

\begin{figure}
\begin{center}
\includegraphics[scale=0.3]{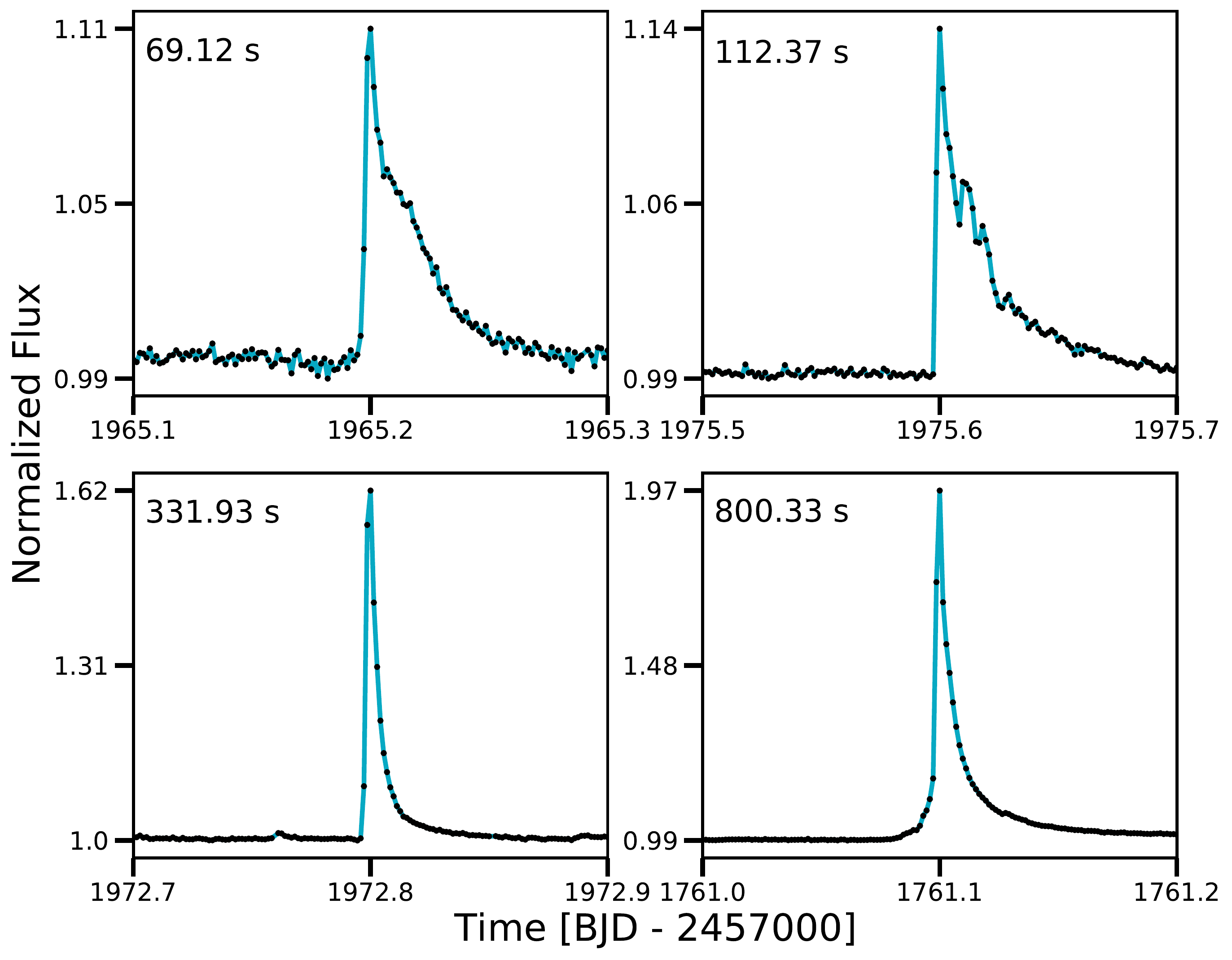}
 \caption{{Four most energetic optical flares observed in the TESS data, with the calculated equivalent duration for each noted in the upper-left. Note the different $y$-scale for each panel. The brightest flare nearly doubles the observed flux in the TESS bandpass.}}
\label{fig:brightest_flares}
\end{center}
\end{figure}

The flares in this sample cover an energy range of $2.44 \times 10^{30} - 3.66 \times 10^{35}$\,erg, with a median energy of $1.31\times10^{31}$\,erg. We highlight the four brightest flares observed by TESS from CR\,Dra in Figure~\ref{fig:brightest_flares}. Their equivalent durations and energies (assuming grey isotropy and scaling from a Gaia luminosity of $0.121 L_\odot$, not accounting for binarity) are, in ascending order: $69.12$\,s and $3.10 \times 10^{33}$\,erg; $112.37$\,s and $7.07 \times 10^{33}$\,erg; $331.93$\,s and $9.40 \times 10^{34}$\,erg; $800.33$\,s and $3.66 \times 10^{35}$\,erg. As the Gaia luminosities do not account for binarity, these reported energies are indicative, and likely accurate only to within a factor of $\sim 2$.

We detect a high flare rate of $2.30$ flares per day, which is towards the upper end of the \citet{Gunther2019} M dwarf TESS flare sample. Since TESS does not resolve CR\,Dra, this flare rate is for the binary. The flare rate was calculated by weighting each flare by the probability assigned by \texttt{stella}. Such a high flare rate in the 600-1000\,nm TESS bandpass is unsurprising for such an active M dwarf system as CR Dra, and consistent with the $B$-band flare rate of $\approx$2.4 flares per day \citep{2018JAVSO..46...21V}. 

We measure a consistent rotation period of 1.984$\pm$0.003 days across each sector of data using \texttt{stella} and the Lomb-Scargle periodogram \citep{lomb76,scargle82} as implemented in \textsc{Astropy} \citep{2013A&A...558A..33A,jvp}. There is no evidence for any other significant period in the TESS lightcurve. 

We can also use this rotation period to infer the inclination of the rotational axis of the component whose rotation is detected by TESS using the hierarchical Bayesian inference code \texttt{inclinations} \citep{white_pleiades}. We adopt the CARMENES spectroscopic rotational velocity measurement of  $v\,\sin{i} = 17.36\pm0.55$\,km\,s$^{-1}$ from \citet{jeffers18}, an effective temperature of 4128\,K \citep{deka18}, and from the effective temperature a radius of $0.56\pm0.05\text{R}_\odot$ using the relation of \citet{cassisi19} and nominal $\sim 10\%$ uncertainty.

We can propagate these uncertainties using hierarchical Bayesian inference in \textsc{PyStan} \citep{stan}. This leads to strong constraints indicating a high rotational inclination: at the $99.9^\text{th}$ percentile, $i>62^\circ$, with a median posterior sample of $i = 83^\circ$. The posterior is largely flat between 70 and 90$^\circ$. 

This inclination constraint may be spurious: using the point estimate parameters, $\sin\,{i} \sim 1.21$, and to get these inclinations we are stretching the uncertainties on radius in particular by at least $0.05\text{R}_\odot$, or $v\sin\,{i}$ is overestimated. As CR~Dra is a binary, the CARMENES reported spectroscopic parameters \citep{jeffers18} are subject to more systematic than statistical uncertainty. While these spectroscopic parameters are the best available, they have have been determined without considering binarity. The measurement may be accurate for one component or an average of both, and without access to the raw data and pipeline it is not straightforward to quantify this effect. The CARMENES estimated spectral type is M1.0, and the differential photometry from interferometric observations by \citet{2008AJ....136..974T} suggest component spectral types of M0 and M3. 

However, the interferometric orbit derived by \citet{2008AJ....136..974T} is inconsistent with the separate radial-velocity orbits determined by either \citet{sperauskas19} (1.57 yrs) or \citet{shkolnik2010} (530\,d). We are not confident in identifying which orbit solution is correct and suggest that if the two stars are spectrally similar, the $v\,\sin{i}$ measurements from rotational broadening will be dominated by the more rapidly-rotating star.

We believe that the flaring component in both optical and radio emission is likely from the more rapidly rotating component, the same star giving rise to the rotational modulation, as fast rotators are have previously found to be bright flare stars \citep{2012ApJ...746...23M,feinstein20}. We do not detect evidence of a second rotational modulation curve in the light curve, so either the other star is inactive or rotating with a very similar period. The latter has been rarely observed \citep{feinstein20}.

Since both components are M~dwarfs of similar mass, these derived stellar parameters should be reasonable. It is therefore plausible that we are observing the more active star at a high inclination, even equator-on, though to confirm this we would want to obtain better spectroscopic parameters. Such an inclination is consistent with the high radio detection rate if the radio emission is similar to auroral emission observed on Jupiter, as discussed in detail in Section\,\ref{sec:discuss}. It will be important in future work to use detailed models of spectroscopic data, together with further interferometric observations, to disentangle the spectral types of the two stars and identify the flaring component(s). For the purposes of this paper, it is sufficient to note that at least one component is brightly flaring, with a clearly identified rotation period, and that the other appears significantly more quiescent.

Furthermore, since the CR\,Dra is located at 20.4\,pc \citep{2018A&A...616A...1G}, even the smallest proposed orbit by \citet{shkolnik2010} implies that the binary is wide enough that no sub-Alfv\'{e}nic interaction is occurring between the two stars \citep{Vedantham2020}.

\section{LOFAR observations and data reduction}
\label{sec:lofar_data}

The LOFAR data were taken in two sets of runs separated by approximately a year -- May to July 2014 and June to August 2015. Due to the desire to observe the ELAIS-N1 field at a high elevation relative to LOFAR, as to minimise the impact of the ionosphere, each observation was taken at a similar local sidereal time (LST). CR\,Dra was located $\approx$0.8 degrees away from the pointing centre for each epoch of observation, corresponding to approximately the 80\% power point of the primary beam at 146\,MHz. We excluded the LOFAR observation L346136 (2015-06-14) due to poor quality data at the location of CR\,Dra, likely produced by problematic ionospheric solutions near CR\,Dra. In total, we have 156.3\,h ($\approx$6.5\,days) of data on CR\,Dra. The observing information and local rms noise for each of the 21 LOFAR observations of CR\,Dra are listed in Table\,\ref{tab:radio_observations}. 

\begin{table}
  \small
  \caption{\label{tab:radio_observations} Observing information for the LOFAR observations of CR\,Dra. `Start time' and `Length' correspond to the time an observation started in UTC and the total duration of the observation, respectively. $\sigma_{I}$ and $\sigma_{V}$ represent the local rms noise near CR\,Dra in Stokes I and V for an image synthesised over the entire duration and bandwidth of the corresponding observation, respectively. The date of the observations are reported in YYYY-MM-DD format.}
  \begin{center}
    \begin{tabular}{lccccc}
      \hline
      LOFAR ID & Date & Start time & Length & $\sigma_{I}$ & $\sigma_{V}$  \\
       & &  & (h) & ($\mu$Jy) & ($\mu$Jy) \\
       \hline
       \hline 
      L229064 & 2014-05-19 & 19:49:19 & 8.0 & 108 & 92\\
      L229312 & 2014-05-20 & 19:46:23 & 8.0 & 92 & 81\\
      L229387 & 2014-05-22 & 19:30:00 & 8.0 & 98 & 76\\
      L229673 & 2014-05-26 & 19:30:00 & 8.0 & 101 & 70\\
      L230461 & 2014-06-02 & 19:30:00 & 8.0 & 100 & 85\\
      L230779 & 2014-06-03 & 19:30:00 & 8.0 & 99 & 75\\
      L231211 & 2014-06-05 & 19:30:00 & 8.0 & 98 & 75\\
      L231505 & 2014-06-10 & 19:50:00 & 7.3 & 109 & 80\\
      L231647 & 2014-06-12 & 19:50:00 & 7.0 & 105& 79\\
      L232981 & 2014-06-27 & 20:05:58 & 5.0 & 117 & 92\\
      L233804 & 2014-07-06 & 19:59:00 & 5.0 & 137& 111\\
      L345624 & 2015-06-07 & 20:11:00 & 7.7 & 115& 81\\
      L346154 & 2015-06-12 & 20:11:00 & 7.7 & 280& 148\\
      L346454 & 2015-06-17 & 20:11:15 & 7.7 & 320& 185\\
      L347030 & 2015-06-19 & 17:58:00 & 7.7 & 108& 78\\
      L347494 & 2015-06-26 & 20:11:00 & 7.7 & 262& 111\\
      L347512 & 2015-06-29 & 20:11:00 & 7.7 & 121& 87\\
      L348512 & 2015-07-01 & 20:11:00 & 6.7 & 160& 112\\
      L366792 & 2015-08-07 & 18:11:00 & 7.7 & 136& 101\\
      L369548 & 2015-08-21 & 16:11:00 & 7.7 & 105& 81\\
      L369530 & 2015-08-22 & 16:11:00 & 7.7 & 110& 91\\
    \hline
  \end{tabular}
\end{center}
\end{table}

The calibration and data reduction strategies performed on the LOFAR data are identical to those outlined by \citet{Pepe2020}, \citet{Tasse2020}, and described briefly in Section 5.1 of \citet{2019A&A...622A...1S}. All pointings had a native 1\,s integration time and 13.0\,kHz spectral resolution covering 114.9 to 177.4\,MHz. After running the standard LoTSS deep field pipeline on each epoch, all sources outside of a 10$'$ radius region around CR\,Dra were removed using the direction dependent calibration solutions. The datasets were phase shifted to the position of CR\,Dra, while accounting for LOFAR station beam attenuation. \textsc{DPPP} \citep{2018ascl.soft04003V} was used to solve residual phase errors by applying a self-calibration loop on 10-20\,s timescales. This total electron content and phase self-calibration loop was followed by several rounds of diagonal gain calibration on timescales of $\approx$20\,min on the phase-corrected data (van Weeren et al. in prep.). The self-calibration timescales are determined by several bright compact sources in a 10$'$ region around CR\,Dra that have peak flux densities $\gtrsim$0.1\,Jy. Bad data are rejected based on the diagonal gain solutions flagging large outliers. During this self-calibration loop automatic clean masking was employed.

Stokes I and V images of each epoch of CR\,Dra were then made from these reduced datasets using a robust parameter \citep{1995AAS...18711202B} of $-$0.5 via \textsc{WSClean} \citep[v\,2.6.3;][]{2014MNRAS.444..606O}. For the time series analysis, each image was synthesised over all of the available 62.5\,MHz bandwidth to maximise the signal. The flux density scale for each image was set using the flux density scale of the deep image \citep{Pepe2020}, whose flux density scale was established by the calibrator 87GB\,J160333.2+573543. Some of the observations in 2015 had poor ionospheric conditions, resulting in $\sim$1.5 times higher noise than the observations conducted in 2014 (see Table\,\ref{tab:radio_observations}).

\subsection{Producing the radio lightcurve}

Our sampling function of the radio data was determined by the Stokes I noise $\sigma_{I}$ in each 8 hour integrated epoch, as listed in Table\,\ref{tab:radio_observations}. We decided to scale our binning by the noise of the 8\,h observation as ECMI emission can have variable maser conditions that can result in non-detections in uniformly-binned data, complicating our periodicity search. Adaptive binning provides unnecessary details at this stage of the analysis since we are searching for a near 2 day signal. We split a LOFAR observation evenly into three or two time intervals if $\sigma_{I} \leq 110$\,$\mu$Jy or $110$\,$\mu$Jy $< \sigma_{I} \leq 160$\,$\mu$Jy, respectively. If $\sigma_{I} > 160$\,$\mu$Jy, only one image was made for the epoch. Such time sampling was a compromise between maximising signal-to-noise in all epochs and remaining sensitive to a potential signal of the photometric rotation period of CR\,Dra (as discussed further in Section\,\ref{sec:flares_prot}). We did not change the sampling for the Stokes V in order to allow an accurate calculation of the fraction of circularly polarised emission. Additionally, such sampling of the radio data produces a well-defined window function for the time series analysis. 

The flux density of CR\,Dra in each image was measured using the Background And Noise Estimator (\textsc{BANE}) and source finder \textsc{Aegean} \citep[v\,2.1.1;][]{2012MNRAS.422.1812H,2018PASA...35...11H}. To correctly account for uncertainties associated with non-detections in our time series analysis, we used the priorised fitting option of \textsc{Aegean} at the location of CR\,Dra. We fitted for the shape and flux density of CR\,Dra as the effective point spread function is influenced by ionospheric conditions. Such a scheme is similar to forced photometry fitting in optical astronomy when variable seeing conditions apply. In the Stokes V images we searched for both significant positive and negative emission. 

Similar to \citet{CallinghamPopulation}, we define the sign of the Stokes V emission as left-hand circularly-polarised light minus right-hand circularly-polarised light. Therefore, a positive Stokes V measurement implies the detected light is more left-hand polarised than right-hand polarised. This definition is followed in the pulsar community but is the reverse of the IAU convention \citep{2010PASA...27..104V}. 

\subsection{Producing dynamic spectra of the radio bursts}

For the bright bursts detected in the LOFAR observations conducted on 2014-05-19, 2014-06-02, and 2015-06-07, we constructed dynamic spectra from image space by synthesising 300 images of 3.125\,MHz bandwidth and 0.53\,h duration for each observation. Similar to above, the flux density of CR\,Dra in each image was measured via priorised fitting using \textsc{BANE} and \textsc{Aegean}. The dynamic spectra were then formed from these measured flux densities. The two spectral channels centred around $\approx$150\,MHz for the 2014-06-02 epoch were discarded due to intense radio frequency interference (RFI). Forming dynamic spectra in image space allows for reliable identification of real emission when it is of low significance.

Finally, the burst detected in the 2014-06-02 epoch was so bright it was possible to form a dynamic spectra for all the Stokes parameters directly from the visibilities. We did this using the LOFAR package \texttt{DynSpecMS}\footnote{\url{https://github.com/cyriltasse/DynSpecMS}} (Tasse et al. in prep), which allows us to examine the time and frequency dependence of the residual data at a specific pixel position using natural weighting of the visibilities. The dynamic spectra have a time and spectral resolution of 8.05\,sec and 78.1\,kHz, respectively. The instrumental leakage between Stokes I and the other Stokes parameters is $<2\%$ \citep{2019A&A...622A..16O}, as also confirmed by inspecting the dynamic spectra of other non-polarised sources in the field.

\section{Radio lightcurve and time series analysis}

\subsection{Radio lightcurve}

\begin{figure*}
\begin{center}
\includegraphics*[scale=0.6]{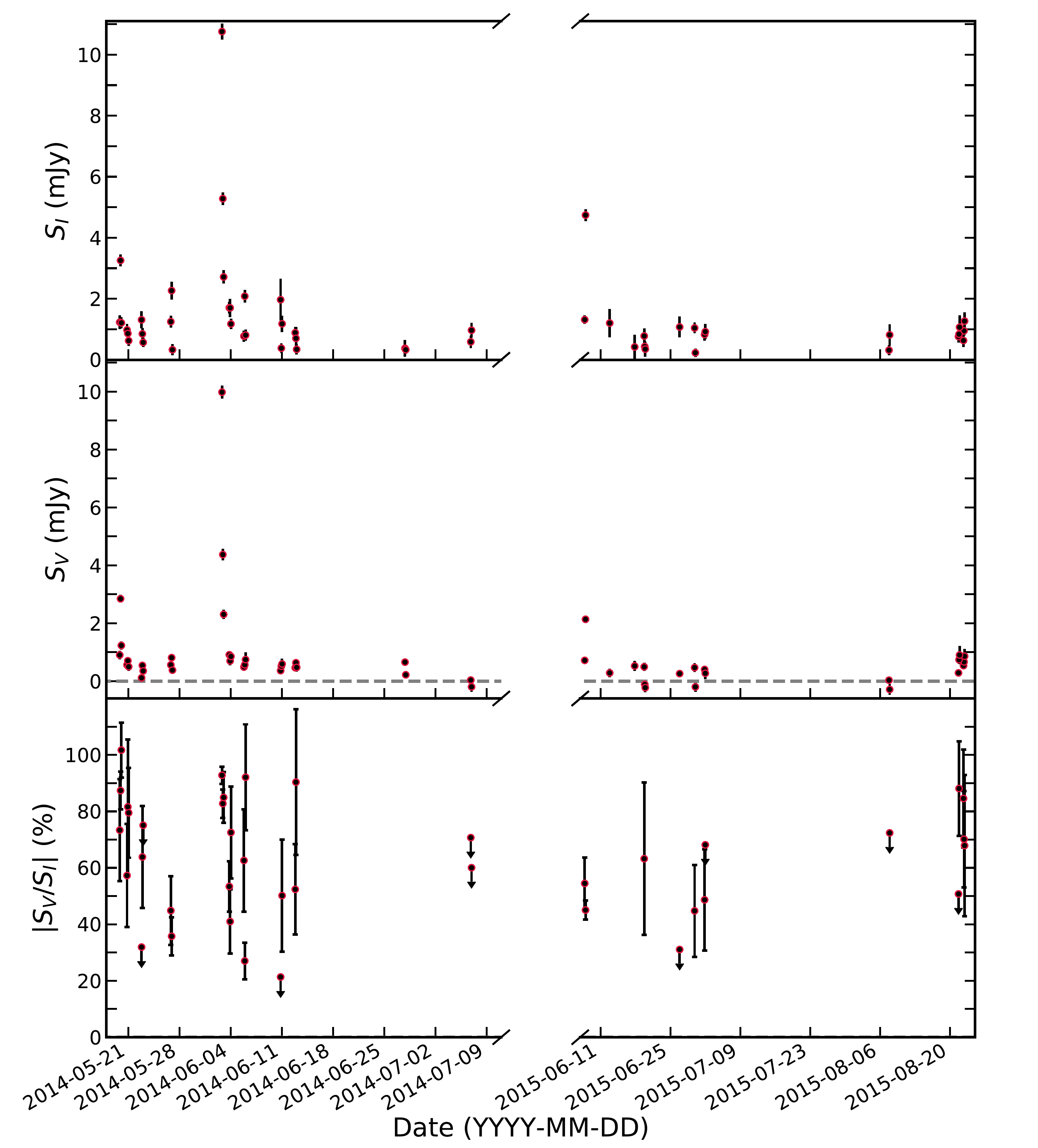}
 \caption{Longterm 146\,MHz radio lightcurve of CR\,Dra in total intensity $S_{I}$ (top panel), circular polarisation $S_{V}$ (middle panel), and the ratio $|S_{V} / S_{I}|$. Only positive Stokes V emission is detected, implying we observe only left-hand circularly polarised emission from CR\,Dra. Uncertainties represent 1$\sigma$ and are only shown if larger than the symbol size. The ratio $|S_{V} / S_{I}|$ is plotted if the corresponding signal-to-noise ratio of CR\,Dra is $\geq3\sigma$ in both Stokes I and Stokes V emission. We show 3$\sigma$ upperlimits for $|S_{V} / S_{I}|$ if the emission in Stokes I is $\geq3\sigma$ but $<3\sigma$ in Stokes V.}
\label{fig:longterm_lc}
\end{center}
\end{figure*}

We present the 146\,MHz Stokes I $S_{I}$ and V $S_{V}$ lightcurve for all of the available data on CR\,Dra in Figure\,\ref{fig:longterm_lc}. The circularly polarised fraction $|S_{V} / S_{I}|$ for an observation is only plotted if CR\,Dra is detected with a signal-to-noise ratio $\sigma \geq 3$ in both Stokes I and V emission. A 3$\sigma$ upperlimit for $|S_{V} / S_{I}|$ is shown if the Stokes I emission from CR\,Dra is $\geq3\sigma$ but the Stokes V emission is $<3\sigma$. The values plotted in Figure\,\ref{fig:longterm_lc} are also provided in Table\,\ref{tab:cr_flux}.

As is shown in Figure\,\ref{fig:longterm_lc}, CR\,Dra is detected in total intensity at a significance $\geq 3\sigma$ in all of our observations, except for the those conducted on 2015-06-12 and 2015-06-17. These two observations had the poorest ionospheric conditions of the monitoring campaign, resulting in a local rms noise $\sim$3 times larger than average. If we conservatively consider the June 2015 observations as non-detections, when a noise floor of $\approx$0.1\,mJy is reached, we detect 146\,MHz emission from CR\,Dra at $\geq$3$\sigma$ 90$^{+5}_{-8}$\% of the time. The reported Wilson interval on the detection rate corresponds to 1$\sigma$. The median and semi-interquartile-range (SIQR) of the total intensity emission from CR\,Dra in our observations is 0.92$\pm$0.31\,mJy.

The low-frequency emission we have detected from CR\,Dra is extremely variable. CR\,Dra displays at least a factor of two variation in total intensity emission within two-thirds of the observations that we split into two or three time intervals. Additionally, while $\approx$90\% of the total intensity emission from CR\,Dra is $<$\,2.2\,mJy, we observed CR\,Dra to burst to flux densities $\approx$1.6 to 5.3 times brighter in three different epochs (2014-05-19, 2014-06-02, and 2015-06-07). By far the largest of these bursts was detected in the 2014-06-02 epoch, reaching a $\approx$2.7\,h band-averaged flux density of 10.75$\pm$0.26\,mJy. Theses bursts are discussed further below in Sections\,\ref{sec:dyn_spec} and \ref{sec:weaker_flares}. The bursts detected in these three epochs are the only emission exceeding three times the median flux density, implying we have a chance of 14$^{+9}_{-6}$\% of detecting such bursts in an observation. This is calculated assuming the bursts are stochastically driven, implying no dependence on the rotational phase of the CR\,Dra. 

The circularly polarised emission from CR\,Dra also displays significant variability, largely tracing the variations in total intensity. We detect $\geq3\sigma$ circularly polarised emission in at least a portion of all the observations with $\geq3\sigma$ Stokes I emission, expect for the observations conducted on 2014-07-06, 2015-08-07, and 2015-06-26. For the first two epochs, this is consistent with $|S_{V} / S_{I}| \lesssim 60\%$. For the 2015-06-26 epoch, a non-detection in Stokes V implies $|S_{V} / S_{I}| \lesssim 30\%$.

$|S_{V} / S_{I}|$ has a median of 66\% and a large SIQR of 33\% for the observations in which we detect both Stokes I and V emission. Such a wide variation in the fraction of circularly polarised light is evident in the bottom panel of Figure\,\ref{fig:longterm_lc}. The fraction of circularly polarised emission varies from $\approx90\%$ for the bursts observed on 2014-05-19 and 2014-06-02 to 27$\pm$6\% and lower for a portion of the 2014-06-05 epoch. When CR\,Dra is not bursting, the circularly polarised fraction can also be quite variable within an epoch. For example, in the 2014-06-05 epoch $|S_{V} / S_{I}|$ varies from 27$\pm$6\% to 92$\pm$19\% within 2.5\,h.

\subsection{Radio time series analysis}

We want to ascertain whether the detected radio emission is related to the 1.984\,d rotation period present in the TESS lightcurve. We searched for radio periodicity in several different datasets produced by various cuts to the LOFAR data: filtering out bursts (epochs with Stokes $I$ flux densities $>1.5$\,mJy, flux density uncertainties $>8.5$\,mJy, or too close in time to the major flare during the 2014-06-02 epoch), separately on data from 2014 and 2015, and finally on all the available radio data. 

We applied the Lomb-Scargle periodogram in its \textsc{Astropy} implementation, following a similar procedure to that outlined in Section\,\ref{sec:flares_prot}. We did not find any significant power for any of the datasets outlined above that could not be explained by the window function. This does not rule out a relationship with the stellar rotation since, unfortunately, the LOFAR observations were all taken at roughly similar LSTs. This implies a 1.984\,d optical rotation period could be hidden by the window function of the radio observations. The interpretation of a non-detection of radio periodicity in light of the breakdown of co-rotation and star-planet interaction models will be discussed further in Section \ref{sec:discuss}. 

Finally, we note that the two brightest coherent bursts we detect are only separated by 14$^{\circ}$ of phase, accurately known since they were observed only two weeks apart. The only other significant burst we detect occurs $\approx$180$^{\circ}$ offset in phase relative to these two bursts. With such bursts being over three times brighter than the median detected flux density, and offset by $\approx$180$^{\circ}$ in phase, this has an intriguing resemblance to the satellite-driven beamed radio emission on Jupiter \citep{marques2017,2018A&A...618A..84Z}. This comparison will be explored further in Section\,\ref{sec:discuss}.

\subsection{Dynamic spectrum of the 2014-06-02 epoch}
\label{sec:dyn_spec}

\begin{figure}
\begin{center}
\includegraphics[scale=0.35]{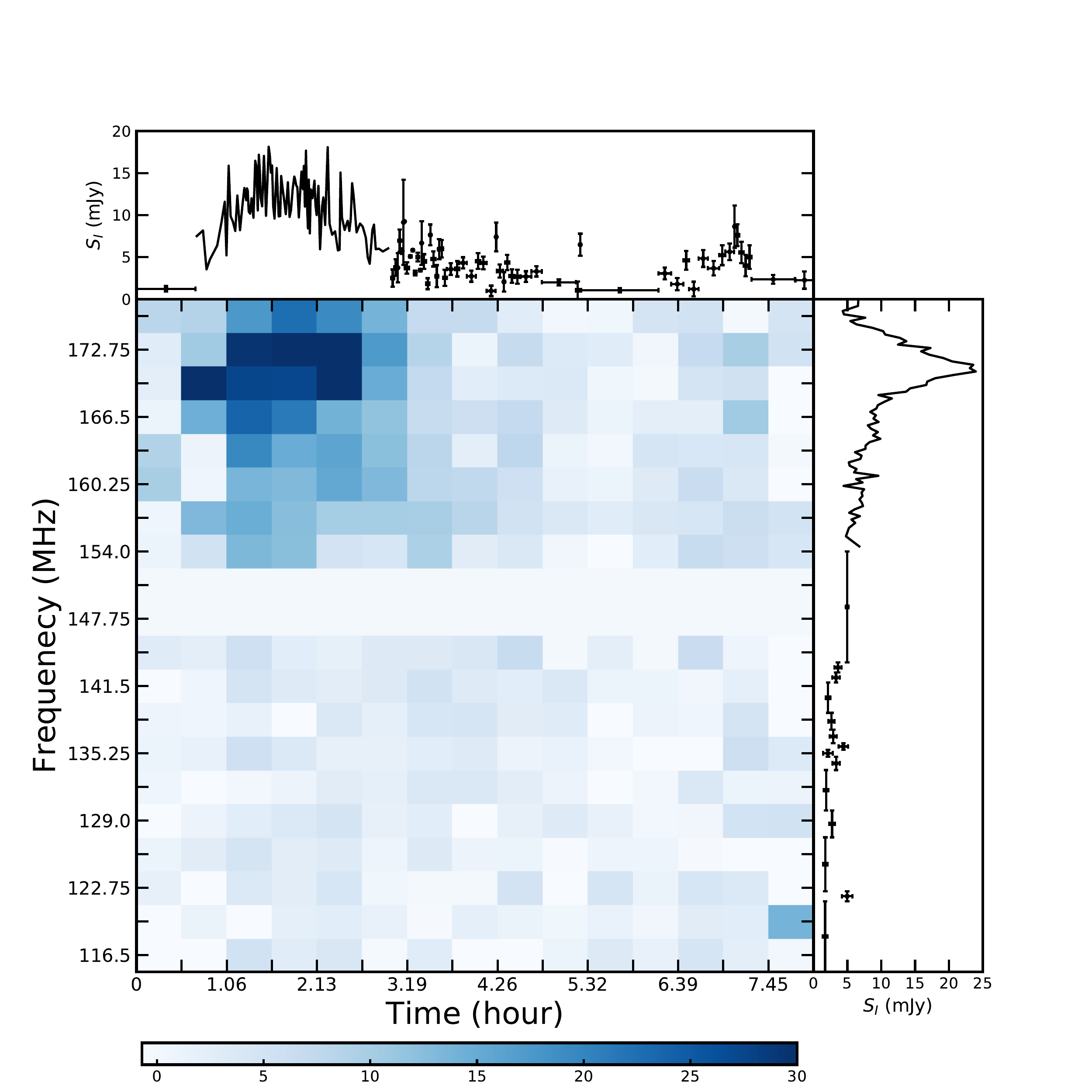}
 \caption{Dynamic spectrum formed from images of CR\,Dra for the 2014-06-02 epoch in total intensity. The median noise in each image used to form this dynamic spectrum is 1.7\,mJy, with $\approx$25\% of the pixels corresponding to a $\geq3 \sigma$ detection. The burst in the upper-left saturates at $\approx$40\,mJy. The colour bar at the bottom communicates the flux density scale in mJy. The white bar centred around $\approx$150\,MHz corresponds to the frequency range for which radio frequency interference prevented reliable data being recovered. The top and righthand panels are the lightcurve and spectrum of CR\,Dra if integrating completely over bandwidth and duration, respectively. For high signal-to-noise areas, the flux density is represented with a line that has uncertainty to better than 10\%. These two side plots show that CR\,Dra is detected at all times and at all frequencies. }
\label{fig:dyn_spec}
\end{center}
\end{figure}

We investigated the time and frequency structure of the bright burst detected in the 2014-06-02 epoch by constructing the dynamic spectrum presented in Figure\,\ref{fig:dyn_spec} from image space (see Section\,\ref{sec:lofar_data}). The dynamic spectrum shows a significant burst that lasts $\approx$2.5\,h, and is largely confined to a bandwidth of 12.5\,MHz centred on $\approx$170\,MHz. The accompanying side panels in Figure\,\ref{fig:dyn_spec} demonstrate that CR\,Dra is also detected at all times and frequencies of the epoch. While the burst dominates for $\approx$2.5\,h, the system relaxes to a level of 3$\pm$1\,mJy for the rest of the observation. There is a hint of a second burst at $\approx$7\,h into the observation but at a much lower significance. The frequency structure between 116 to 155\,MHz is also largely flat with an average flux density of 3$\pm$1\,mJy.  

\begin{figure*}
\begin{center}
\includegraphics[scale=0.5]{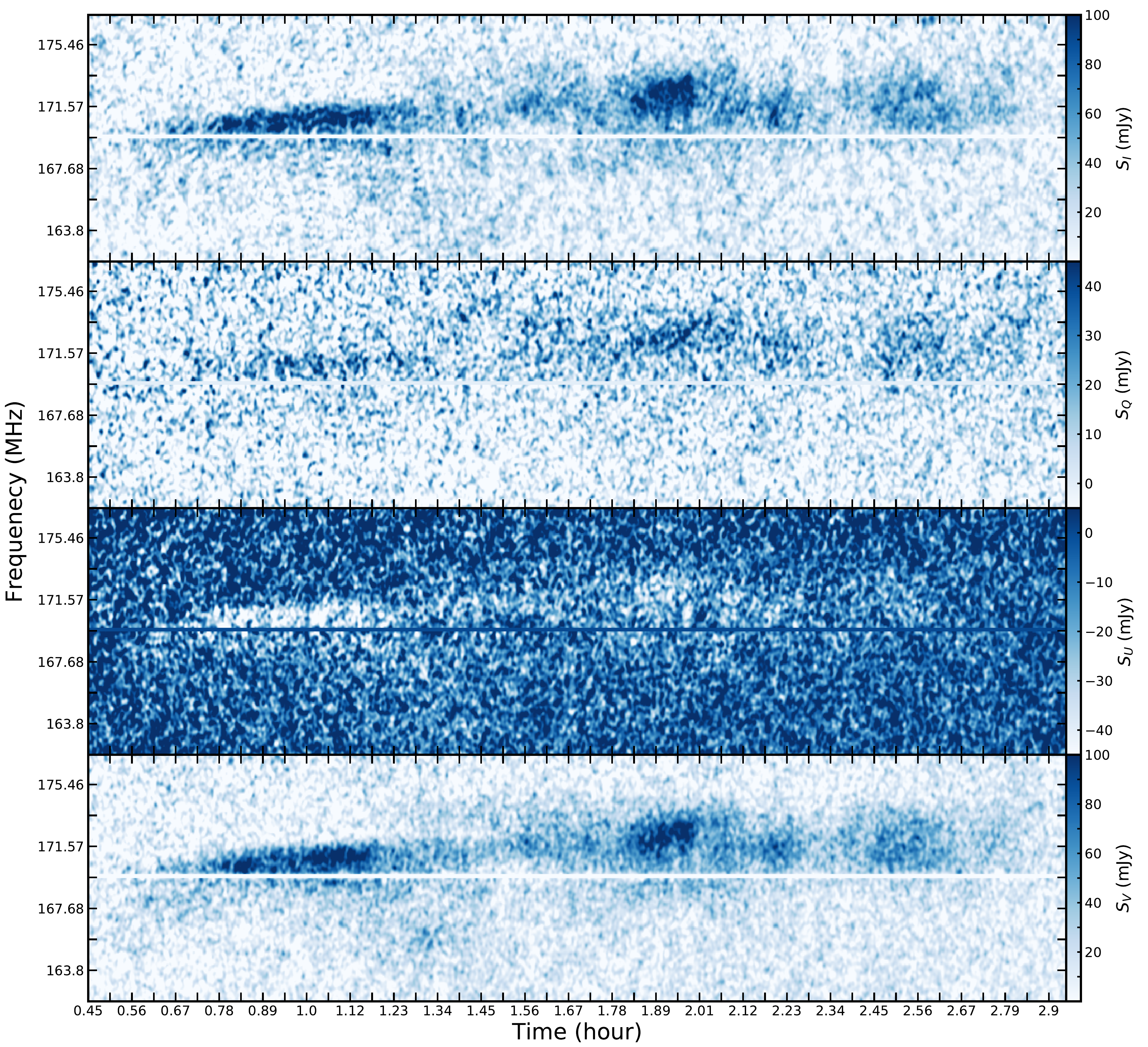}
 \caption{Dynamic spectra of the burst from CR\,Dra detected during the 2014-06-02 observation in Stokes I (first panel from top), Q (second panel), U (third panel), and V (bottom panel) formed directly from calibrated visibilities. The time and frequency resolution of the dynamic spectra are 8.05\,sec and 78.1\,kHz. A Gaussian filter with a standard deviation of 1.5 pixels has been applied. The burst saturates at $\approx$205, $\approx$80, $\approx$-90, and $\approx$198\,mJy, in Stokes I, Q, U, and V. The detection of emission from the burst in Stokes Q and U implies the emission is elliptically polarised at least part of the time. The rms noise in the Stokes I dynamic spectrum is 16\,mJy, while it is 12\,mJy for the Stokes Q, U, and V dynamic spectra. The colour bars on the right communicate the flux density scale for each Stokes parameter. The thin white/dark blue bar centred on $\approx$169.7\,MHz represents bandwidth that has been excised due to RFI.}
\label{fig:dyn_spec_flare}
\end{center}
\end{figure*}

To analyse the burst in greater detail, we present a dynamic spectrum for all Stokes parameters in Figure\,\ref{fig:dyn_spec_flare} which is formed directly from the calibrated visibilities. The burst saturates at $\approx$205\,mJy in Stokes I. Such a burst corresponds to an isotropic radio spectral luminosity $L_{\nu}$ of 1.0$\times 10^{17}$\,ergs\,s$^{-1}$\,Hz$^{-1}$. All of the emission in the phase space around the burst is $\gtrsim90\%$ in Stokes V. The detection of emission in Stokes Q and U demonstrates that the burst is elliptically polarised. The burst is highly localised to $\approx$4\,MHz of bandwidth, starting at 170.1\,MHz and drifting to 173.9\,MHz with time. The burst has a frequency drift of $\approx$3.1\,MHz\,h$^{-1}$, and appears to widen in frequency with time.

Such a bright coherent burst resembles the 154\,MHz bursts observed on UV\,Ceti \citep{2017ApJ...836L..30L} but it is a factor of 2 more luminous and appears to last a factor of three longer. Similarly, the burst is also much more confined in frequency space, longer in duration, and/or one to three orders of magnitude more spectrally luminous than the majority of the bursts observed on AD\,Leo, UV\,Ceti, and EQ\,Peg at frequencies $\gtrsim325$\,MHz \citep{2019ApJ...871..214V}.

\begin{figure}
\begin{center}$
\begin{array}{cc}
\includegraphics[scale=0.35]{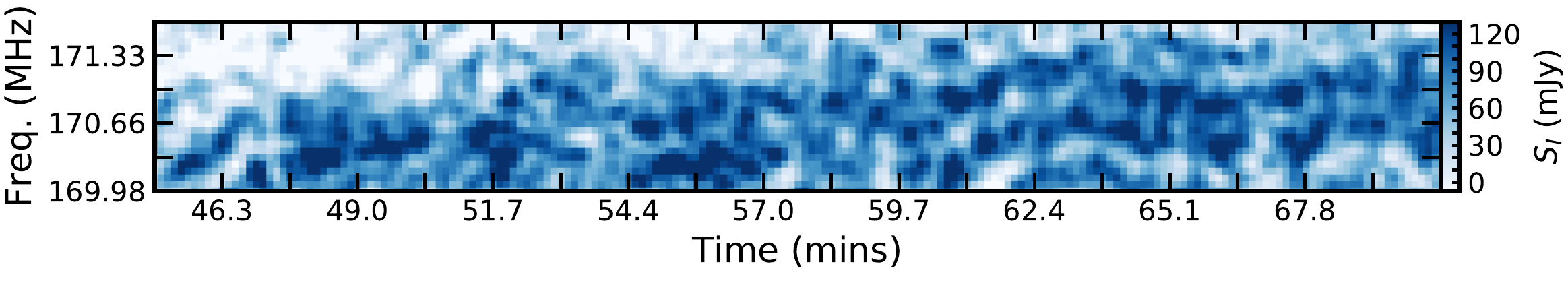} \\
\includegraphics[scale=0.35]{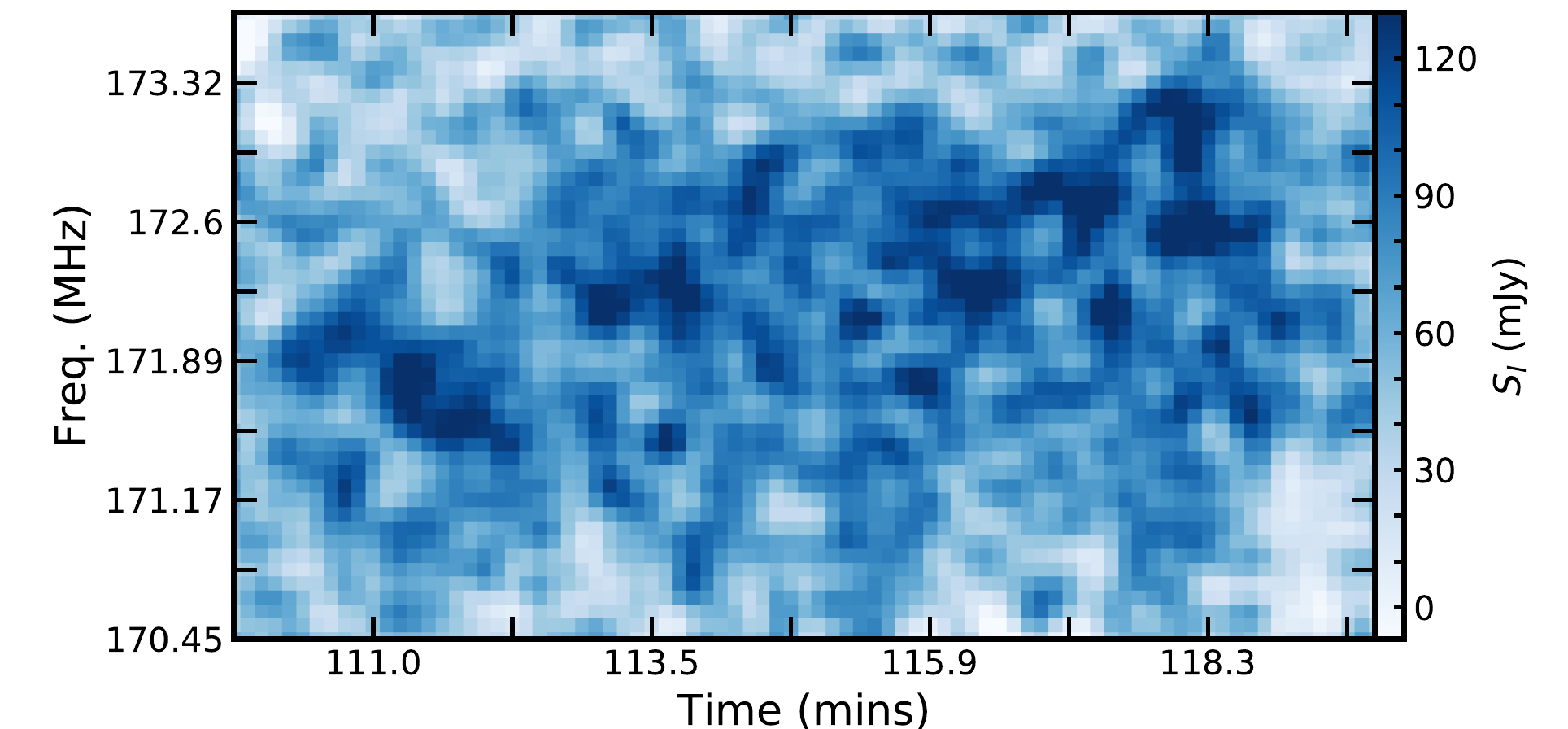}\\
\end{array}$
 \caption{A close up of total intensity emission for two different areas in the bright burst presented in Figure\,\ref{fig:dyn_spec_flare}. We have used a slightly different flux density scale than that used in Figure\,\ref{fig:dyn_spec_flare} to emphasise the mottled structure and sub-bursts. A Gaussian filter with a standard deviation of 1.0 pixel has been applied. The rms noise is 16\,mJy.}
\label{fig:mottled_dynspec}
\end{center}
\end{figure}

The burst also appears to have a mottled structure, with unresolved sub-bursts present in the broad structure. We provide two close ups of the burst in Figure\,\ref{fig:mottled_dynspec} to highlight the sub-burst structure. Such structures also persist if we change the interferometric weighting scheme from natural to uniform. These sub-bursts have a frequency-time slope of $\lesssim 0.15$\,MHz\,s$^{-1}$, with breaks between sub-bursts lasting between 8 and $\approx$24\,sec when the 2.5\,h long broad burst is most active. The sub-burst structures are reminiscent of the long (L)-bursts seen from Jupiter \citep{1983phjm.book..226C,Ellis1974}, which have a drift rate of $\lesssim 0.1$\,MHz\,s$^{-1}$ and modulation lanes of reduced emission that last $\lesssim$1\,min \citep{1970A&A.....4..180R,1978Ap&SS..56..503R}. L-bursts from Jupiter can be produced by both the interaction with Io and the breakdown of rigid co-rotation in the Jovian magnetosphere \citep{1998JGR...10320159Z} and previously observed for brown dwarfs \citep[e.g.][]{2008ApJ...684..644H}. Moreover, the top panel of Figure\,\ref{fig:mottled_dynspec} appears to show two parallel narrow bands of emission undulating in phase. Such a banded signal has previously been observed in Jupiter's decametric radio emission \citep{2018A&A...610A..69P}, reinforcing the similarity of the emission from CR\,Dra to that of Jupiter.

\subsection{Dynamic spectra of the 2014-05-19 and 2015-06-07 epochs}
\label{sec:weaker_flares}

\begin{figure*}
\begin{center}$
\begin{array}{cc}
\includegraphics[scale=0.35]{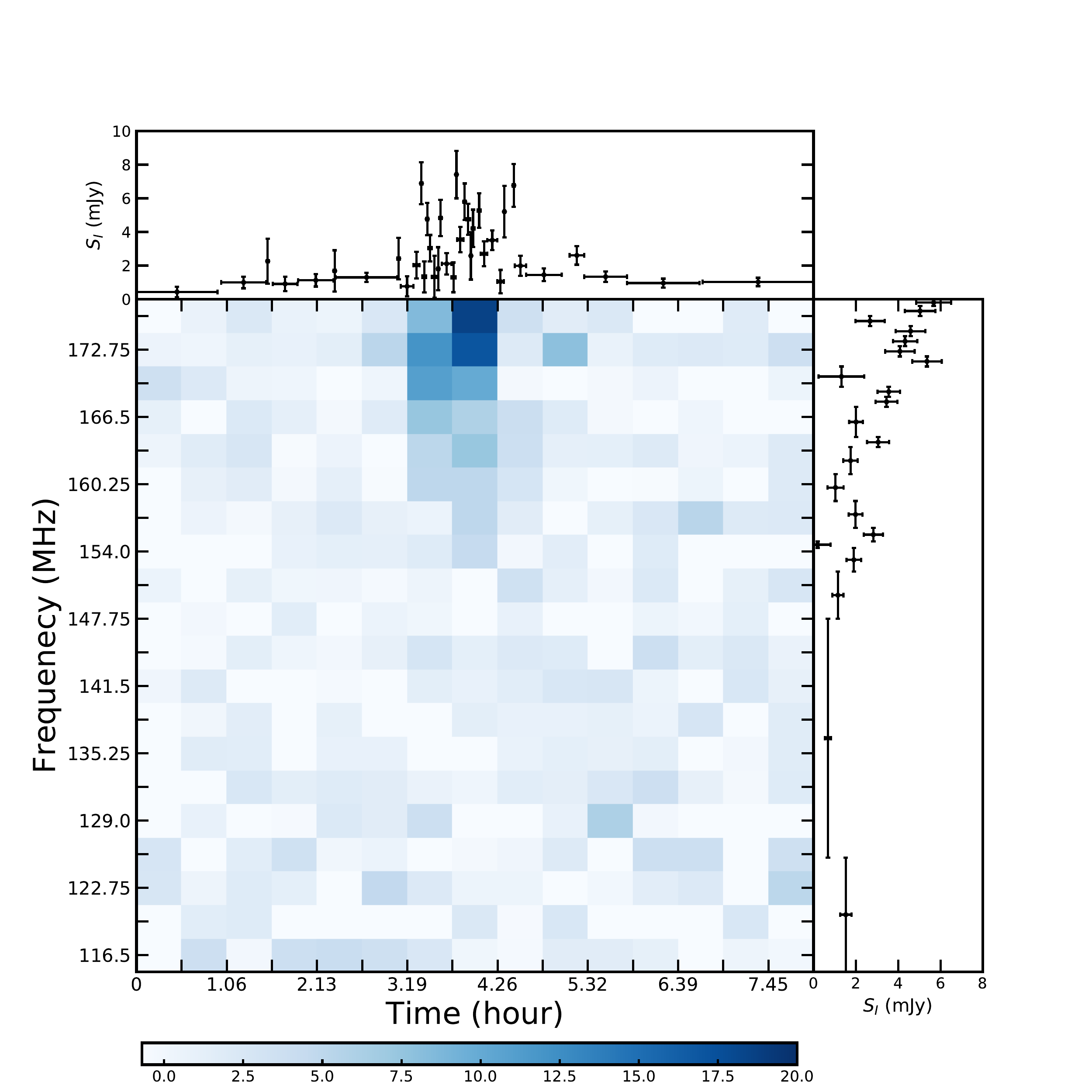} &
\includegraphics[scale=0.35]{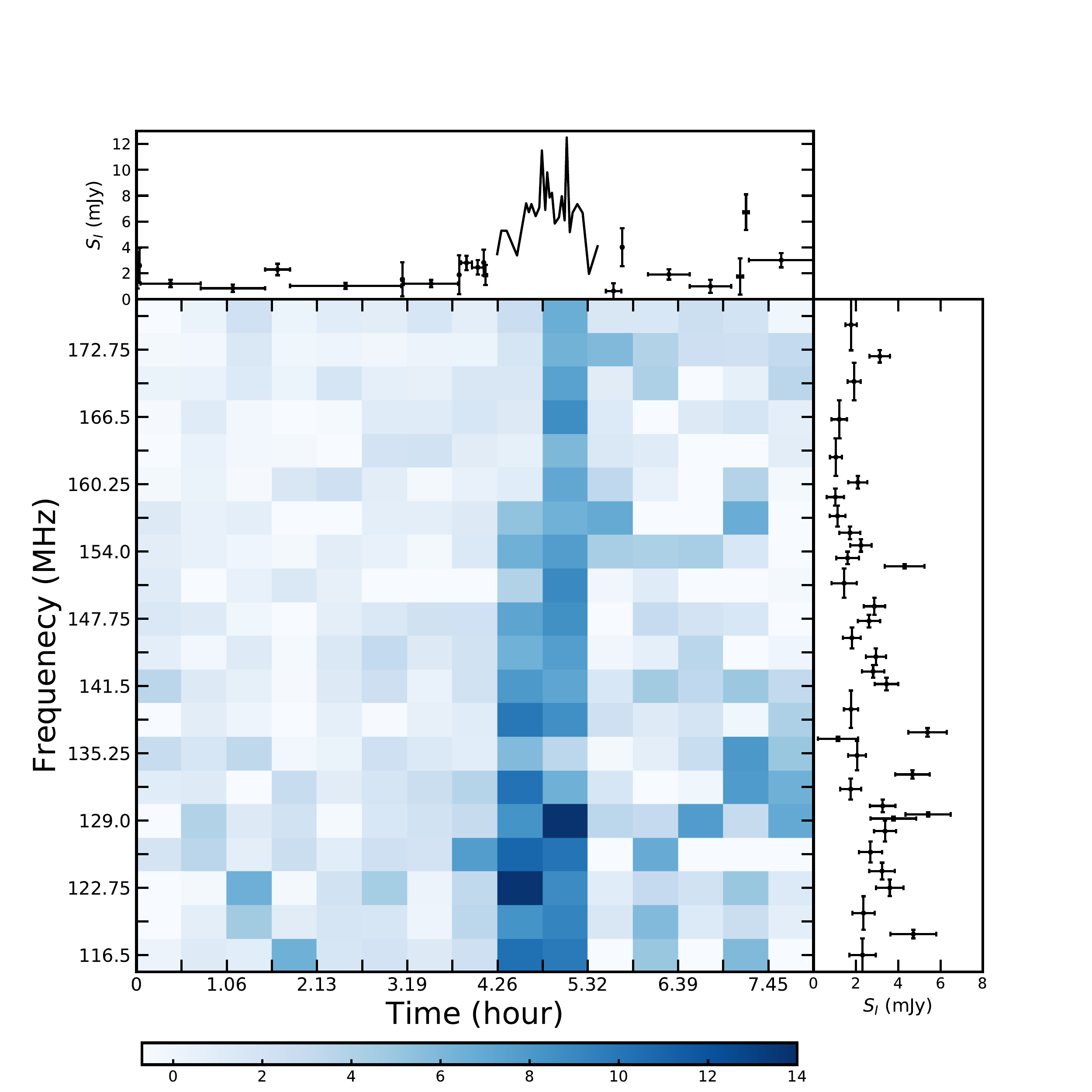}\\
\end{array}$
 \caption{Total-intensity dynamic spectra of the bursts detected from CR\,Dra in the 2014-05-19 (left panel) and 2015-06-07 (right panel) epochs formed in image space. The median noise for both epochs is $\approx$1.5\,mJy. The bursts reach $\approx$19 and $\approx$14\,mJy for the 2014-05-19 and 2015-06-07 epochs, respectively. The colour bar at the bottom of each panel communicates the flux density scale in mJy for the corresponding observation. The top and righthand plots on each pseudo-dynamic spectrum are the lightcurve and spectrum of CR\,Dra if integrating completely over bandwidth and duration, respectively. For high signal-to-noise areas, the flux density is represented with a line that has uncertainty to better than 10\%. The side plots show that CR\,Dra is detected at all times and at all frequencies for both epochs.}
\label{fig:weaker_flares}
\end{center}
\end{figure*}

We also present the dynamic spectra, constructed from image space, of the significant bursts detected during the 2014-05-19 and 2015-06-07 epochs in Figure\,\ref{fig:weaker_flares}.

The burst detected during the 2014-05-19 observation shows similar characteristics to the burst shown in Figure\,\ref{fig:dyn_spec} -- namely it appears highly-confined in frequency space around 173\,MHz and is $>90\%$ circularly polarised. However, it should be noted that the burst also could extend to frequencies higher than available with our observations. Additionally, the burst only lasts $\approx$1.1\,h. 

In comparison, the burst detected during the 2015-06-07 is broadband, detected at all frequencies from 116 to 178\,MHz. The burst also appears to march up in frequency, with no emission detected at frequencies $>160$\,MHz for the first half of the burst. The burst has a frequency drift rate of $\approx$12\,MHz\,h$^{-1}$, before completely dissipating within our bandwidth after $\approx$1.1\,h. The burst is $>$70\% circularly polarised when we can detect at least a 3$\sigma$ source in total intensity.

Finally, we note that we cannot provide the true dynamic spectra for these two bursts. For the 2014-05-19 epoch, problematic RFI conditions make interpretation of the burst in finer detail than presented in Figure\,\ref{fig:weaker_flares} not possible. For the 2015-06-07 epoch, the burst is of too low signal-to-noise and broad band.
 
\section{Discussion}
\label{sec:discuss}

The interpretation of the radio lightcurve, dynamic spectrum, and time series analysis of CR\,Dra is predicated on what we determine to be the emission mechanism: plasma radiation or ECMI emission. If the former is occurring, the detected emission should share similarities to Type\,I, II, III, or IV bursts observed on the Sun \citep{1985ARA&A..23..169D}. If the latter, the emission characteristics will have more in common with auroral processes associated with gas giant planets \citep{1998JGR...10320159Z,2007P&SS...55..598Z} and brown dwarfs \citep{2008ApJ...684..644H,2017ApJ...846...75P,2018haex.bookE.171W}.

There are three properties that can be derived from our observations that allow us to differentiate between the competing mechanisms: brightness temperature, degree and consistency of circular polarisation, and time-frequency structure of the emission. 

The low-frequency and high flux density of the detected emission implies a high brightness temperature for the radiation. If we assume the entire photospheric surface of radius 0.56\,R$_\odot$ is the emission site \citep{cassisi19}, the median detected flux density of 0.9\,mJy corresponds to a brightness temperature of 1.7$ \times 10^{12}$\,K. This is a lower limit as either plasma or ECMI emission is likely confined to much smaller areas, as seen on Jupiter \citep{2004JGRA..109.9S15Z} and the Sun \citep{1985ARA&A..23..169D}. Furthermore, the radius used in this calculation is likely as overestimate as it is derived from an effective temperature measurement convolving both stars in CR\,Dra.

In the following discussion, we assume that the radio emission is produced by only one of the stars in CR\,Dra binary, unless explicitly otherwise stated. We believe such an assumption is valid because: (1) there are many known M dwarf binaries that have stars of similar masses and ages, such as UV\,Ceti \citep{Zic2019}, GJ\,412 \citep{CallinghamPopulation}, and Ross\,867/868 \citep{quiroga20}, but only one of the stars emits in the radio, and; (2) we only see one rotational signal in the TESS lightcurve. In such cases, only the fast rotating star is active in the radio \citep{2012ApJ...746...23M}. Finally, outside of the periodicity search, our analysis is ambivalent to which star is radio-emitting.

\subsection{Plasma radio emission}
\label{sec:plasma}

Can plasma radiation reach a brightness temperature of 1.7$ \times 10^{12}$\,K at 146\,MHz? Since CR\,Dra has a hot corona of $\sim$7\,MK \citep{2015A&A...578A.129J,CallinghamPopulation}, implying a coronal plasma density of at least 10$^{10.5}\,\text{cm}^{-3}$, it is possible for fundamental plasma emission to reach such a brightness temperature at low frequencies. The most common way to generate plasma emission is via an impulsive injection of heated plasma into the corona \citep{1985ARA&A..23..169D,2001A&A...374.1072S}. Assuming a stable hydrostatic density of the corona \citep{Vedantham2020,CallinghamPopulation}, and applying Equations 15 and 22 of \citet{2001A&A...374.1072S}, the brightness temperature reaches $\approx$0.4$ \times 10^{12}$\,K if the impulsive event injected into the corona is 20 times the coronal temperature. The assumptions required in such a calculation are accurate to an order of magnitude as, for example, the coronal plasma density is an upper-limit since CR\,Dra is not resolved in the X-ray observations \citep{2016A&A...588A.103B}. Therefore, it is plausible fundamental plasma emission could produce the observed brightness temperature of the emission. 

However, the consistent sign of circular polarisation suggests that plasma emission is unlikely the driver of the low-frequency radiation. Left-handedness is measured for the circularly-polarised emission in all $\approx$6.5 days of monitoring data, which includes a year separation between the two main blocks of observations. This indicates the radio emission is consistently emerging from a source region with the same magnetic polarity. If the emission were driven by flares, the polarity would be expected to flip since flares occur randomly spread over the stellar disk \citep{1985ARA&A..23..169D}, not solely from regions with identical magnetic field polarity \citep{2019ApJ...871..214V}.

The detection of elliptical polarisation of the bright burst presented in Figure\,\ref{fig:dyn_spec_flare} also suggests that plasma emission is unlikely producing the low-frequency emission \citep{2017ApJ...836L..30L,Zic2019}. For Solar radio plasma emission, any linear polarisation is obliterated by differential Faraday rotation
of the plane of polarisation during passage through the corona \citep[e.g.][]{1985srph.book..289S}. A similar situation is expected to arise in the magnetosphere of the radio-bright star in CR\,Dra, where the scale height is higher than the Sun's due to its large X-ray luminosity. Furthermore, the detection of linear polarisation in other active M\,dwarfs has been used as evidence that plasma emission is not the mechanism producing the radio emission \citep[e.g.][]{Zic2019}.

We note that while it is plausible that plasma emission is generating emission for the detections of CR\,Dra in which the circularly polarised fraction is low, we show below that the observed fluctuations in the circularly polarised fraction can also be readily justified within an ECMI model. Therefore, due to the consistent handedness of the circular polarisation and detected elliptical polarisation of the bright burst, we suggest it is unlikely we are detecting plasma emission from CR\,Dra.

\subsection{ECMI radio emission}

As outlined above, we suggest that the observed highly circularly polarised radiation with brightness temperatures $>10^{12}$\,K is more likely to be generated by ECMI emission. The requirements for ECMI emission to occur are: (1) the cyclotron frequency $\nu_{c}$ to be much less than the plasma frequency $\nu_{p}$, (2) the presence of mildly relativistic electrons, and (3) a population inversion in perpendicular velocity of the mildly relativistic electrons \citep{1985ARA&A..23..169D}. The condition $\nu_{c} / \nu_{p} \ll 1$  can be achieved at high magnetic latitudes for rapidly rotating bodies by the presence of a strong magnetic field, or by a parallel electric field that produces plasma density cavities \citep{1985ARA&A..23..169D,1998JGR...10320159Z}. The necessary population inversion requires some anisotropy in the distribution of energetic electrons. This anisotropy can be introduced by the loss cone of upgoing electrons reflected by the magnetic mirror force, trapped electrons, or parallel acceleration by parallel electric fields, followed by adiabatic evolution of the electron distribution function \citep{Melrose1978}. 

There are three main astrophysical situations that can produce the conditions listed above on a star that could power the maser: a flaring coronal loop, breakdown of co-rotation, and a sub-Alfv\'{e}nic star-planet interaction. 

In a coronal loop, it is possible to set up an unstable loss-cone distribution of electron energies to drive the ECMI emission \citep{1985ARA&A..23..169D,2016A&A...589L...8M,Vedantham2020}. Injection of heated plasma into the loop, combined with magnetic mirroring, sets up the loss-cone distribution. In this case, the brightness temperature is directly related to the size and width of the coronal loop \citep{1985ARA&A..23..169D}. Following the model outlined in \citet{2020NatAs.tmp...34V}, to reproduce the brightness temperature implied by our median detected flux density requires a coronal loop with a length two orders of magnitude greater than the stellar radius of CR\,Dra. Such a large structure is not dynamically stable or previously observed, with the largest coronal loop detected extending only up to $\approx$4 stellar radii \citep{Benz1998}. The coronal loop model also has difficultly explaining the consistent polarity of the Stokes V emission and high duty ratio of CR\,Dra since coronal loops are transient structures that often form in areas of complex magnetic geometry \citep{2019ApJ...871..214V}. 

We explore the possibility that the radio characteristics of CR\,Dra could be modelled by ECMI emission generated by the breakdown of co-rotation, a direct analogue to the process occurring in the Jovian magnetosphere \citep{cowley2001}. Since we have evidence from the TESS data that a star in CR\,Dra is rapidly rotating at 1.984\,d, a coupling current between the star and its magnetosphere can be produced when plasma at $\gtrsim10$ stellar radii lags behind the magnetic field of the star \citep{2011MNRAS.414.2125N,2012ApJ...760...59N,turnpenney17}. The resulting current then accelerates electrons into the corona and chromosphere of the star, setting up the population inversion necessary for ECMI to operate. The resulting ECMI emission is beamed along the edges of a cone at the electron cyclotron frequency, with a brightness temperature $>10^{12}$\,K and high circularly polarised fraction. While in the Jovian system the plasma disk is supplied by the volcanic activity of Io \citep{bagenal13,bagenal17}, we suggest that the high flare rate of CR\,Dra implies the radio-bright star is continually dumping plasma into its magnetosphere. 

However, if a plasma disk is required to exist to drive the radio emission from the radio-bright star in CR\,Dra, it is not clear how it is sustained when the corona is propelling a stellar wind. A similar problem has been encountered when trying to explain coherent radio emission from massive stars via a direct application of the Jovian breakdown of co-rotation model \citep[e.g.][]{das2019,leto2020}. Potentially, a strong magnetic field could channel the wind is such way to maintain a plasma disk within the stellar magnetosphere of a fast-rotating star \citep{maheswaran2009}, but magneto-hydrodyamic (MHD) simulations and a map of the magnetic field topology are required to test this hypothesis. Regardless of the exact driving mechanism, the observed ECMI emission from CR\,Dra shares strong similarities to the auroral emission observed from bodies in the Solar System. 

The high detection rate of 90\% of emission from CR\,Dra can also be circumstantially explained via an auroral model. For Jupiter, its main auroral oval is an axisymmetric annulus of emission that is $\approx$1$^{\circ}$ wide and offset $\approx$\,15$^{\circ}$ from the magnetic pole \citep{2001P&SS...49.1159P,2009JGRA..114.6210N}. The non-Io decametric (non-Io-DAM) emission connected to this auroral oval is beamed at large angles relative to the magnetic field lines at high invariant latitudes. Therefore, for a preferential line of sight, it is possible to always observe the auroral radio emission from one hemisphere since it is also continuously operating \citep{cecconi12,marques2017}. In this model, as one emission site rotates out of the line of sight due to the radiation being beamed and the emission site being connected to the field line of the star, another site moves into the observer's view. The optical data we have on CR\,Dra suggests we are viewing the system at inclinations $\gtrsim 60^{\circ}$ relative to the pole. Such equator-on inclinations support the high detection rate of CR\,Dra, as a more pole-on orientation would mean the emission would be regularly beamed away from our line of sight. 

It is important to note that the non-Io-DAM emission has a strong variation with respect to the longitude of Jupiter due to interactions with the Solar wind \citep{2010A&A...519A..84E,2012P&SS...70..114H}. We would not expect to see this dependency in the emission from the radio-bright star in CR\,Dra as the radiation originates on the star itself, negating any dawn/dusk effects. The partner star is also likely too far away to have a significant impact on the dynamics of the radio-bright star's magnetosphere. Instead, we expect to observe stochastic variations due to changes in the density and temperature of the plasma in the magnetosphere of the star, which enhances or diminishes the auroral-producing current. Such a situation is similar to the observed change in Io-DAM radio emission due to variations in the volcanic activity of Io \citep{2013GeoRL..40..671Y}. 

The auroral model is also supported by the consistent handedness of the circularly polarised light. The constant positive handedness of the circularly polarised light over 6.5 days of monitoring, taken in two observing blocks separated by a year, requires a stable magnetic field arrangement at the emission site. We infer that the electron acceleration is likely occurring at a distance from the radio-bright star where the large-field structure dominates, before accelerating the electrons into the high corona where the radio emission is produced. Such a configuration is readily established at the poles, and implies the emission we detect is dominated by emission sites in one of the hemispheres of the star. 

ECMI emission is expected to be 100\% circularly or elliptically polarised \citep{1985ARA&A..23..169D} but we detect a large variation in circularly polarised fraction in the emission from CR\,Dra, with a median detection of 66\% and a SIQR of 33\%. This variation in the polarisation fraction can be explained by two features in the ECMI produced by breakdown of co-rotation auroral model -- an emission site also being observed from the opposite pole and propagation effects. 

For the former, there is again precedence in the Jovian system, where right-hand emission dominates when observed from Earth due to the stronger magnetic field at northern Jovian latitudes \citep{1983phjm.book....1A,2010A&A...519A..84E}. Therefore, it is possible that we sometimes observe emission sites from the opposite pole on the radio-bright star in CR\,Dra but the emission is never bright enough to dominate the emission from the pole in which we have a preferential line of sight.

In regards to propagation effects, the circular polarised fraction of emission can be reduced by dispersion or scattering in the coronal plasma \citep{1991A&A...245..299G}, reflections off boundary layers of various density ratios \citep{2006ApJ...637.1113M}, and strong mode coupling in quasi-transverse magnetic field regions \citep{1998ARA&A..36..131B,lamy11}. It is possible all three effects are underway in the Jovian system \citep{1983phjm.book..226C,1988pre2.conf..299L}, with the average and SIQR circularly polarised fraction of non-Io emission from the Northern Hemisphere being 40$\pm$20\%, when averaged over a similar timescale as our data \citep[see e.g. Figs. 8 and 9;][]{marques2017}. Since the radio-bright star in CR\,Dra still possesses a corona, as opposed to Jupiter and ultra-cool dwarfs \citep{2010ApJ...709..332B,2014ApJ...785....9W}, it seems reasonable to expect that different coronal conditions will cause fluctuations in the circularly polarised fraction of the emission. 

Finally, we note that while a sub-Alfv\'{e}nic star-planet interaction can also generate the observed radiation characteristics. While we would have expected a periodicity in the radio light curve \citep{2020NatAs.tmp...34V,CallinghamPopulation}, our time sampling means we can not rule this possibility out. We discuss this model in more detail in reference to the observed radio bursts in the following section.

\subsection{Implications of resolved, coherent radio bursts}

In Sections\,\ref{sec:dyn_spec} and \ref{sec:weaker_flares} we presented dynamic spectra that resolved three of the brightest radio bursts we detected from CR\,Dra. We can use the polarisation, duration, and time-frequency structures of these bursts to infer the dominant stellar pole of which the emission sites occupy, and whether the emission is consistent within the proposed breakdown of co-rotation auroral model.  

For the brightest burst we detect (Figure\,\ref{fig:dyn_spec}), we are able to show that the burst is confined to $\approx$4\,MHz of bandwidth and lasts $\approx$2.5\,h. Such time-frequency structure is in direct contrast to the expected $\approx$5-30 minute time scale and broadband nature expected from plasma emission associated with a CME \citep{2018ApJ...856...39C}. The extraordinarily high brightness temperature lower limit of $>$3.2$ \times 10^{14}$\,K and elliptical polarisation of the burst conclusively demonstrates that ECMI is the emission mechanism, as outlined in Section\,\ref{sec:plasma}. 

The conclusion that the burst is generated by ECMI emission is also supported by the sub-burst structures unresolved in Figures\,\ref{fig:dyn_spec_flare} and \ref{fig:mottled_dynspec}. Such sub-bursts share a striking resemblance to unresolved L-bursts seen on Jupiter \citep[see e.g. Figs. 3 and 14;][]{marques2017}. The varied time scales and gaps between L-bursts could be due to radiation scattering on the interplanetary plasma or plasma in the corona of the radio-bright star in CR\,Dra. Alternatively, such time scales could reflect the intrinsic variation in the maser power in a variable coronal plasma and magnetic field. Assuming we are detecting fundamental ECMI emission, which generally has the fastest growth rate, the small frequency range of the emission implies it is emerging from a local magnetic strength of 61 to 62\,G. The degree of elliptical polarisation, if intrinsic, also implies the ECMI emission angle relative to the magnetic field is quite large at $63\pm5^{\circ}$ for the brightest parts of the burst \citep{1993P&SS...41..333M,1994A&A...286..683D}. Such an opening angle is similar to that derived from ECMI emission from UV\,Ceti \citep{Zic2019}.

If the observed frequency drift of $\approx$3.1\,MHz\,h$^{-1}$ is due to rotational modulation of beamed emission from a very active structure as it rotates in and out of our view in a dipole magnetic field, it shares similarities to the vertex structures seen in Jovian emission \citep{1998JGR...10320159Z}. These vertex structures in dynamic spectra result from the beaming pattern, polarisation, and rotation of the active radio region \citep{2014P&SS...99..136H}. Applying the physics of vertex structures in low-frequency Jovian emission to the properties of broad burst resolved in Figure\,\ref{fig:dyn_spec_flare} \citep{marques2017}, such as the left-handedness of the circular polarised light and direction of its frequency drift, the emission site we observed for this burst was in the southern magnetic pole and eastern limb of the star as it rotates counter-clockwise relative to its spin axis. Such a conclusion is independent of magnetic field topology.

If the burst occurs 10$^{\circ}$ from the rotation axis of the star that we are viewing equator-on implies, we can estimate the size $l$ of the emission site from the duration of the burst \citep{2017ApJ...836L..30L} using $l = \Delta t v$, where $v$ is the velocity of the emission region and $\Delta t$ is the duration over which we observe the emission. We estimate the emission site size is $\approx$2$\times$10$^{4}$\,km. Such a size is an order of magnitude larger than the main-spot emission site produced by the Jupiter-Io interaction, but similar to the total spot size if the tail is accounted \citep{2002Natur.415.1000G,2013P&SS...88...64B}. Depending on the exact location of the emission site, we can vary the size by a factor of two but not easily by an order of magnitude. Therefore, this implies the burst reaches extraordinary brightness temperatures of $\approx1 \times 10^{16}$ to $2.2 \times10^{17}$\,K for over 2.5 hours. The simple loss-cone maser configuration can not reach such high brightness temperatures \citep{1982ApJ...259..844M}. An efficient maser configuration, such as a horse-shoe maser configuration \citep{2013SSRv..178..695B}, is required to produce the properties of this burst.

\citet{1991A&A...249..250M} demonstrated that elliptically polarised radio emission produced by ECMI implies an extremely low electron density $n_{e}$ in the emission region, namely:

\begin{equation}
    n_{e} \lesssim \alpha(\nu / 25\,\mathrm{MHz}),
    \label{eqn:ne}
\end{equation}

\noindent where $\nu$ is the observing frequency and $\alpha$ is a geometric factor of order unity. Using Equation\,\ref{eqn:ne}, the linear polarisation of the bright burst implies it emerged from an emission site with an electron density $\lesssim$\,6\,cm$^{-3}$, indicating the existence of extreme density cavities within the magnetosphere of the radio-bright star in CR\,Dra. Such emission cavities share similarities to the emission regions of Earth’s auroral kilometric radiation \citep{1998JGR...10320159Z}, and have been suggested to exist on other active M dwarfs despite their large coronal densities \citep{2017ApJ...836L..30L,2019ApJ...871..214V,Zic2019,Zic2020}.

The other two bursts presented in Figure\,\ref{sec:weaker_flares} also support that ECMI is the emission mechanism. The burst detected during the 2014-05-19 observation is $>90\%$ circularly polarised with a brightness temperature $>$3.0$ \times 10^{13}$\,K and lasts $\approx$1.1\,h. The source also occurs at a similar frequency as the bright burst presented in Figure\,\ref{fig:dyn_spec}, suggesting a potentially similar emission site location. However, the emission could extend beyond our available bandwidth.

In contrast, the burst detected in the 2015-06-07 is broadband and has a factor of four faster frequency drift rate than the other two resolved bursts. The burst also persists for $\approx$1.1\,h with a high brightness temperature $>$2.2$ \times 10^{13}$\,K and $>70\%$ circularly polarised fraction. While such broadband structure seems to suggest a different emission mechanism than proposed for the other two bursts, similar broadband and hour-long emission structures are also observed in non-Io DAM emission from Jupiter \citep{1998JGR...10320159Z,2014P&SS...99..136H}. The direction of frequency drift of this burst is also in agreement with the bright burst presented in Figure\,\ref{fig:dyn_spec}, consistent with the fact we are preferentially observing the Southern magnetic pole of the radio-bright star in CR\,Dra as it rotates counter-clockwise.

How do these bursts fit within the auroral model when the emission is at least an order of magnitude larger than the median detected flux density? Both Io- and non-Io-DAM emission from the Jovian system can vary by several orders of magnitude, with the former largely influenced by the long term evolution of Io's volcanic activity and magnetic longitude, while the latter is impacted by different Solar wind conditions \citep{2014P&SS...99..136H}. Some bright non-Io-DAM emission events are shown to be associated with `hotspots' of localised precipitations along Jupiter’s main auroral oval \citep{1993JGR....9818779P,2002Natur.415.1000G,2004AnGeo..22.1799N,2013P&SS...88...64B,2020JGRA..12527222D}. We hypothesise that the bright bursts we observe are could be to hotspots in the magnetosphere of CR\,Dra, where an unusually dense and/or hot plasma facilitates highly localised and variable auroral emission. Such anisotropic plasma distribution within the magnetosphere of CR\,Dra seems reasonable to expect due to the high flare activity of the star.

It is also plausible these ECMI bursts could be produced by a putative terrestrial-sized planet in orbit around the star, as Io DAM emission can exceed the intensity of non-Io DAM emission \citep{marques2017}. However, such emission is expected to be periodic. Interestingly, two of the three prominent bursts we detect are close in rotational phase. If an exoplanet is close enough to the star to drive the ECMI emission, it is possible it is tidally locked \citep{2017CeMDA.129..509B}. Alternatively, this grouping in the bursts could have been produced by an active hotspot that existed for at least two weeks in the magnetosphere of the radio-bright star in CR\,Dra, but such an idea is challenged by the fact plasma hotspots in Jupiter's magnetosphere only last several days before being expended \citep{2002Natur.415.1000G,2013P&SS...88...64B}. Unfortunately, the complete rotational phase of CR\,Dra is  poorly sampled in our monitoring campaign, with more complete sampling of the rotational phase required to tell if this grouping of bursts is statistically significant. 

In closing, caution is warranted in the interpretation of the lack of periodicity in the lightcurve. We do not know the magnetic field topology or whether the magnetic axis of the radio-emitting star in CR\,Dra is aligned with its rotation axis. Since it is possible the radio emitting star could have a toroidal or non-axisymmetric poloidal magnetic field configuration \citep{donati08}, it would result in a wider beaming geometry than predicted by the dipolar magnetic field assumed for the discussion above. Zeeman-Doppler imaging \citep[ZDI;][]{2010MNRAS.407.2269M} of CR\,Dra will be important in determining the magnetic field topology and the mechanism that is driving the auroral radio emission.

\section{Conclusions}

We have presented and analysed $\approx$6.5\,days of low-frequency ($\approx$115 to 177\,MHz) monitoring data on the M dwarf flare star binary CR\,Dra. Together with TESS data from three sectors, we conclude:

\begin{itemize}

\item CR\,Dra has a photometric rotation period of 1.984$\pm$0.003\,d and an extremely high optical flare rate of 2.30 flares per day. Available optical data also suggest we are viewing the system at an inclination $>62^{\circ}$, close to equator-on. We assume that the star with the near two day rotation period signal in the TESS lightcurve is the star emitting the radio emission.

\item CR\,Dra has a near-constant low-frequency detection rate of 90$^{+5}_{-8}$\% when a noise floor of $\approx$\,100\,$\mu$Jy is reached, with a median total intensity flux density and SIQR of 0.92$\pm$0.31\,mJy. Left-handed circularly polarised light is consistently detected from CR\,Dra, with the detected circularly polarised fraction having a median and SIQR of 66$\pm$33\%.

\item With a spectral classification of M0 and M3 \citep{2008AJ....136..974T}, CR\,Dra could contain the earliest M-dwarf known to emit coherent radio emission.

\item There is no evidence for periodicity in the low-frequency radio data. However, the LOFAR observations were conducted at similar LSTs, implying the window function could be hiding the signal due to the nearly exactly two day rotation period. The accuracy to which we know the photometric rotation period implies future radio sampling should be conducted within a time span of $\approx$6\,months to prevent large uncertainties in rotational phase. A periodic signal in the radio lightcurve could also be hard to identify if the radio-emitting star in CR\,Dra does not have a dipolar magnetic field topology.

\item The brightest of the three radio bursts we can resolve in dynamic spectra is elliptically polarised and reaches a total intensity of $\approx$\,205\,mJy, implying a brightness temperature $>$3.2$ \times 10^{14}$\,K. This makes the burst one of the most luminous bursts observed from an M dwarf at low frequencies. The burst is highly confined in frequency space, sweeping through only $\approx4$\,MHz of bandwidth in $\approx$2.5\,h. It also displays a mottled structure, which we suggest are unresolved sub-bursts that have a time-frequency slope $\lesssim 0.15$\,MHz\,s$^{-1}$. Such structures resemble coherent low-frequency emission from Jupiter. Furthermore, it is likely the burst emerges from an emission site of $\approx10^{4}$\,km size, indicating an extraordinary brightness temperatures of $\sim$10$^{17}$\,K. Only an efficient maser configuration, such as a horse-shoe maser, can produce such brightness temperatures. 

\item The consistent circularly polarised left-handedness of the detections and elliptical polarisation of the bright burst make it unlikely that plasma emission is the emission mechanism producing the observed low-frequency radiation from CR\,Dra. We suggest that the mechanism that produces the persistent and bursty emission is the ECMI. 

\item The observed ECMI emission shares strong similarities to non-Io-DAM emission from the Jovian system, implying an auroral model best explains the emission properties. However, determining the magnetic field topology of the radio-bright star in CR\,Dra is required to validate if this model is realistic.

\item If the auoral model is applicable, and the radio-bright star in CR\,Dra has a dipolar magnetic field, it is likely have a preferential line of sight to one of the hemispheres of the star. Assuming such emission is similar to the vertex structures seen in Jovian emission, the observed frequency drift direction and left-handedness of the resolved bursts suggests the star is rotating counter-clockwise and the emission we detect is emerging from the Southern magnetic hemisphere.

\end{itemize}

An implication of our study is that CR\,Dra will almost always be detected at low frequencies provided a noise floor of $\approx$\,100$\mu$Jy is reached. Furthermore, ZDI \citep{2010MNRAS.407.2269M} of CR\,Dra would be able to test if the Southern hemisphere of the star is orientated towards us, the direction of stellar rotation, the topology of the magnetic field, and its strength. Due to the fast rotation of one of the stars in CR\,Dra, it is possible that it has a kiloGauss magnetic field \citep{2019A&A...626A..86S}. This means the coherent radio emission from CR\,Dra could extend up to gigahertz frequencies, such as those observed by the Very Large Array Sky Survey \citep[VLASS;][]{2020PASP..132c5001L}.

A MHD model of how the stellar flares deposit plasma in the magnetosphere of the star, and what are the necessary physical conditions of that plasma to maintain the radio emission, would be useful in testing our suggestion that the auroral emission can be sustained through that pathway. Future simultaneous observations of CR\,Dra with TESS and LOFAR would also help test if there is correlation between the optical and radio activity of the star. 

\begin{acknowledgements}
We thank the anonymous referee for a thorough report that led to important improvements to the manuscript. JRC thanks C.~Lynch (Curtin University) for useful discussions, and the Nederlandse Organisatie voor Wetenschappelijk Onderzoek (NWO) for support via the Talent Programme Veni grant. The LOFAR data in this manuscript were (partly) processed by the LOFAR Two-Metre Sky Survey (LoTSS) team. This team made use of the LOFAR direction independent calibration pipeline (\url{https://github.com/lofar-astron/prefactor}), which was deployed by the LOFAR e-infragroup on the Dutch National Grid infrastructure with support of the SURF Co-operative through grants e-infra 160022 e-infra 160152 \citep{2017isgc.confE...2M}. The LoTSS direction dependent calibration and imaging pipeline (\url{http://github.com/mhardcastle/ddf-pipeline/}) was run on compute clusters at Leiden Observatory and the University of Hertfordshire, which are supported by a European Research Council Advanced Grant [NEWCLUSTERS-321271] and the UK Science and Technology Funding Council [ST/P000096/1]. This paper includes data collected by the TESS mission. Funding for the TESS mission is provided by the NASA Explorer Program. This work was performed in part under contract with the Jet Propulsion Laboratory (JPL) funded by NASA through the Sagan Fellowship Program executed by the NASA Exoplanet Science Institute. This work acknowledges the support from the National Science Foundation Graduate Research Fellowship Program under Grant No.\,(DGE-1746045). Any opinions, findings, and conclusions or recommendations expressed in this material are those of the authors and do not necessarily reflect the views of the National Science Foundation. LL was supported by the CNRS/INSU programme of plantelogy. PNB is grateful for support from the UK STFC via grant ST/R000972/1. RJvW acknowledges support from the ERC Starting Grant ClusterWeb 804208. TPR acknowledges support from the European Research Council through Advanced Grant No.\,743029.

This research has made use of the SIMBAD database, operated at CDS, Strasbourg, France, and NASA's Astrophysics Data System. This work has also made use of the \textsc{IPython} package \citep{PER-GRA:2007}; SciPy \citep{scipy}; \textsc{matplotlib}, a \textsc{Python} library for publication quality graphics \citep{Hunter:2007}; \textsc{Astropy}, a community-developed core \textsc{Python} package for astronomy \citep{2013A&A...558A..33A}; \textit{lightkurve}, a Python package for Kepler and TESS data analysis \citep{lightkurve};  \textsc{PyStan}, the \textsc{Python} interface to the probabilistic programming language \textsc{Stan} \citep{stan}; and \textsc{NumPy} \citep{van2011numpy}. 
\end{acknowledgements}

%
%

\bibliographystyle{aa.bst}
\bibliography{cr_dra.bbl}

\begin{appendix}
\onecolumn
\section{LOFAR flux density measurements of CR Dra}

The flux density measurements and epoch information for the LOFAR observations of CR\,Dra that is plotted in Figure\,\ref{fig:longterm_lc}, and used in the periodicity search, are listed in Table\,\ref{tab:cr_flux}.

\begin{table*}[h]
  \caption{\label{tab:cr_flux} Flux density of CR Dra in Stokes I and V for the epochs presented in Figures\,\ref{fig:longterm_lc}. Reported times are in UTC. $S_{I}$ and $S_{V}$ represent the flux density of CR\,Dra in Stokes I and V, respectively. The date of the observations are reported in YYYY-MM-DD format. $|S_{V} / S_{I}|$ is only reported if the signal-to-noise ratio of the emission from CR\,Dra is $\geq3\sigma$ in both Stokes I and Stokes V. 3$\sigma$ upperlimits for $|S_{V} / S_{I}|$ are provided if the emission in Stokes I is $\geq3\sigma$ but $<3\sigma$ in Stokes V. The upperlimits are calculated using 3 times the rms noise in Stokes V.}
  \begin{center}
    \begin{tabular}{lcccc}
      \hline
      Date & Time & $S_{I}$ & $S_{V}$ &  $|S_{V} / S_{I}|$ \\
       &  & (mJy) & (mJy) & (\%) \\
       \hline
       \hline 
2014-05-19 & 19:49:27.0 & 1.23 $\pm$ 0.23 &  0.90 $\pm$ 0.15     & 73$\pm$18 \\
2014-05-19 & 22:29:26.3 & 3.26 $\pm$ 0.19 & 2.85  $\pm$ 0.14    & 87$\pm$7\\
2014-05-20 & 01:09:25.5 & 1.21 $\pm$ 0.18 & 1.23  $\pm$ 0.14     & 101$\pm$10 \\
2014-05-20 & 19:46:31.0 & 0.98 $\pm$ 0.20 & 0.56 $\pm$ 0.14     & 57 $\pm$  18 \\
  2014-05-20 & 22:26:30.3 & 0.86 $\pm$  0.20 & 0.70 $\pm$ 0.19   & 83  $\pm$  24 \\
  2014-05-21 & 01:06:29.5 & 0.63 $\pm$  0.17 & 0.50 $\pm$ 0.14   & 80 $\pm$  16\\
  2014-05-22 & 19:30:08.0 & 1.31 $\pm$     0.28 & 0.11 $\pm$ 0.10   & $<$32 \\
2014-05-22 & 22:10:07.3 & 0.85 $\pm$ 0.19 & 0.54 $\pm$ 0.09     & 64 $\pm$  18\\
2014-05-23 & 00:50:06.5 & 0.57 $\pm$ 0.15& 0.35 $\pm$ 0.14     & $<$75\\
  2014-05-26 & 19:30:08.0 & 1.25 $\pm$ 0.20& 0.56     $\pm$  0.13   & 45 $\pm$  12\\
  2014-05-26 & 22:10:07.3 & 2.27 $\pm$ 0.30 & 0.81         $\pm$ 0.11   & 36 $\pm$   7  \\
  2014-05-27 & 00:50:06.5 & 0.39 $\pm$  0.18 & 0.38 $\pm$ 0.12  & \\
  2014-06-02 & 19:30:08.0 & 10.76 $\pm$ 0.26 & 9.98      $\pm$ 0.23    & 93 $\pm$   3\\
2014-06-02 & 22:10:07.3 & 5.28$\pm$ 0.21 & 4.37  $\pm$ 0.20     & 83 $\pm$   5\\
2014-06-03 & 00:50:06.5 & 2.71 $\pm$ 0.22 & 2.31  $\pm$ 0.16    & 85  $\pm$   9\\
2014-06-03 & 19:30:08.0 & 1.70 $\pm$ 0.19 & 0.91  $\pm$ 0.11     & 53 $\pm$   9\\
2014-06-03 & 22:10:07.3 & 1.70 $\pm$ 0.30 & 0.70      $\pm$ 0.15           & 41   $\pm$  11\\
2014-06-04 & 00:50:06.5 & 1.18 $\pm$  0.17 & 0.85  $\pm$ 0.15    & 73  $\pm$  16\\
  2014-06-05 & 19:30:08.0 & 0.78 $\pm$  0.18 & 0.49 $\pm$ 0.08  & 63 $\pm$  18\\
  2014-06-05 & 22:10:07.3 & 2.08 $\pm$ 0.20 & 0.56    $\pm$ 0.12  & 27  $\pm$   6\\
2014-06-06 & 00:50:06.5 & 0.81 $\pm$ 0.19 & 0.74 $\pm$ 0.25     & 92 $\pm$  19\\
2014-06-10 & 19:50:08.0 & 1.97 $\pm$ 0.68 &  0.36      $\pm$     0.13      & $<$21\\
  2014-06-10 & 22:15:05.3 & 0.38 $\pm$  0.16 & 0.52 $\pm$ 0.11        & \\
2014-06-11 & 00:40:02.6 & 1.18 $\pm$ 0.27 & 0.59 $\pm$ 0.19     & 50  $\pm$  20\\
  2014-06-12 & 19:50:08.0 & 0.89 $\pm$  0.19 & 0.46 $\pm$ 0.10   & 52 $\pm$  16\\
2014-06-12 & 22:09:59.3 & 0.70 $\pm$ 0.23 & 0.63  $\pm$ 0.11    & 90 $\pm$  26\\
  2014-06-13 & 00:30:06.7 & 0.34 $\pm$  0.17 & 0.47 $\pm$ 0.15   & \\
2014-06-27 & 20:06:06.0 & 0.38 $\pm$     0.28 & 0.34  $\pm$ 0.20    & \\
  2014-06-27 & 22:36:09.3 & 0.33 $\pm$  0.13 & 0.22 $\pm$ 0.13  & \\
2014-07-06 & 19:59:08.0 & 0.59 $\pm$ 0.19 & 0.04 $\pm$ 0.12     & $<$71\\
  2014-07-06 & 22:29:12.5 & 0.97 $\pm$ 0.25 & -0.20  $\pm$ 0.18   & $<$60 \\
  2015-06-07 & 20:11:08.0 & 1.31 $\pm$ 0.14 & 0.72     $\pm$ 0.09   & 54 $\pm$   9\\
2015-06-08 & 00:01:03.1 & 4.74 $\pm$  0.20 & 2.14  $\pm$ 0.13     & 45  $\pm$   3\\
2015-06-12 & 20:11:08.0 & 1.20 $\pm$ 0.47 & 0.28  $\pm$ 0.15     & \\
  2015-06-17 & 20:11:23.0 & 0.42 $\pm$  0.40 & 0.52 $\pm$ 0.18   & \\
  2015-06-19 & 17:58:08.0 & 0.78 $\pm$  0.25 &  0.49  $\pm$ 0.14   & 63 $\pm$  27\\
2015-06-19 & 20:31:24.7 & 0.43 $\pm$ 0.14 & -0.12 $\pm$ 0.12     & \\
    2015-06-19 & 23:04:41.5 & 0.35 $\pm$  0.25 & -0.23  $\pm$ 0.16 & \\
2015-06-26 & 20:11:08.0 & 1.07 $\pm$  0.34 & 0.26  $\pm$ 0.11     & $<$31\\
2015-06-29 & 20:11:08.0 & 1.04 $\pm$ 0.18 & 0.47 $\pm$ 0.15     & 44 $\pm$  16\\
    2015-06-29 & 23:47:26.0 & 0.29 $\pm$  0.14 & -0.20  $\pm$ 0.17 & \\
  2015-07-01 & 20:11:08.0 & 0.83 $\pm$  0.20 & 0.40 $\pm$ 0.11   & 49 $\pm$  18\\
    2015-07-01 & 23:31:08.6 & 0.91 $\pm$  0.25 & 0.27      $\pm$ 0.21  & $<$68\\
2015-08-07 & 18:11:08.0 & 0.31 $\pm$ 0.17 & 0.03 $\pm$ 0.10    & \\
2015-08-07 & 21:23:24.0 & 0.82 $\pm$ 0.27 & -0.29 $\pm$ 0.19      & $<$72\\
2015-08-21 & 16:11:08.0 & 0.76 $\pm$ 0.20& 0.28 $\pm$ 0.11    & $<$51 \\
  2015-08-21 & 18:44:24.7 & 0.84 $\pm$  0.27 & 0.73 $\pm$ 0.15   & 87  $\pm$  17\\
2015-08-21 & 21:17:41.5 & 1.07 $\pm$ 0.39 & 0.90        $\pm$  0.31        \\
  2015-08-22 & 16:11:08.0 & 0.63 $\pm$  0.22 & 0.54 $\pm$ 0.12  & 85  $\pm$  17\\
2015-08-22 & 18:44:24.7 & 0.95 $\pm$ 0.16 & 0.67  $\pm$ 0.12    &  70 $\pm$  17\\
  2015-08-22 & 21:17:41.5 & 1.27 $\pm$ 0.29& 0.86    $\pm$ 0.25   & 68 $\pm$   25\\
  \hline
  \end{tabular}
\end{center}
\end{table*}

\end{appendix}
\end{document}